\documentclass[aip,jcp,amsmath,amssymb,floatfix,reprint,citeautoscript]{revtex4-2}

\usepackage[utf8]{inputenc}
\usepackage{graphicx}
\usepackage{wrapfig}
\usepackage{placeins}
\usepackage[colorlinks,allcolors=black,citecolor=blue,urlcolor=blue]{hyperref}

\usepackage{float}

\graphicspath{{figures/}}

\newcommand{\vect}[1]{\mathbf{#1}}

\begin{document}

\def\mytitle{
Committee neural network potentials control generalization errors
and~enable~active~learning
}
\title{\mytitle}

\author{Christoph Schran}
\email{christoph.schran@rub.de}

\author{Krystof Brezina}

\author{Ondrej Marsalek}
\email{ondrej.marsalek@mff.cuni.cz}

\affiliation{Charles University, Faculty of Mathematics and Physics, Ke Karlovu 3, 121 16 Prague 2, Czech Republic}

\date{\today}

\begin{abstract}

\setlength\intextsep{0pt}
\begin{wrapfigure}{r}{0.4\textwidth}
  \hspace{-1.5cm}
  \includegraphics[width=0.4\textwidth]{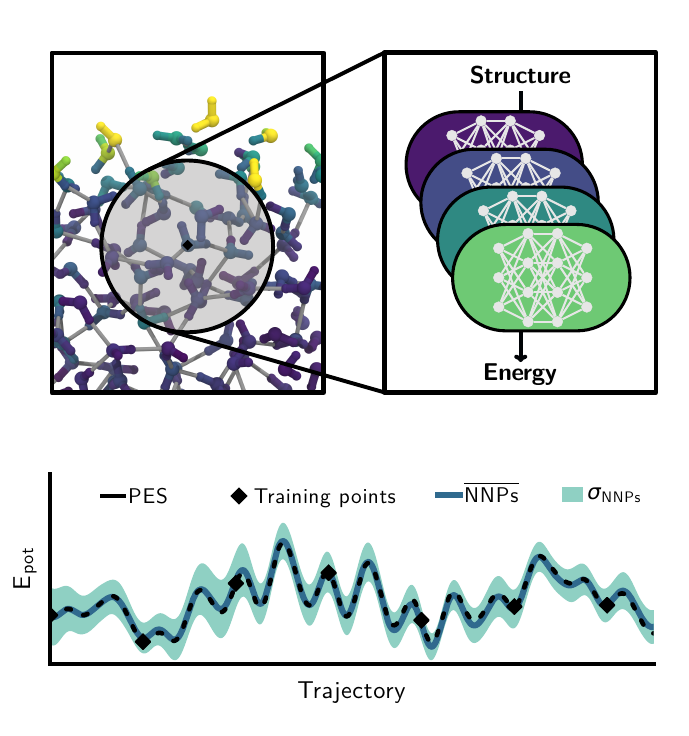}
\end{wrapfigure}

It is well known in the field of machine learning that committee models improve accuracy,
provide generalization error estimates, and enable active learning strategies.
In this work, we adapt these concepts to interatomic potentials based on artificial neural networks.
Instead of a single model, multiple models that share the same atomic environment descriptors yield an average that outperforms its individual members as well as a measure of the generalization error in the form of the committee disagreement.
We not only use this disagreement to identify the most relevant configurations to build up the model's training set in an active learning procedure, but also monitor and bias it during simulations to control the generalization error.
This facilitates the adaptive development of committee neural network potentials and their training sets, while keeping the number of ab initio calculations to a minimum.
To illustrate the benefits of this methodology, we apply it to the development of a committee model for water in the condensed phase.
Starting from a single reference ab initio simulation, we use active learning to expand into new state points and to describe the quantum nature of the nuclei.
The final model, trained on 814 reference calculations, yields excellent results under a range of conditions, from liquid water at ambient and elevated temperatures and pressures to different phases of ice, and the air-water interface --- all including nuclear quantum effects.
This approach to committee models will enable the systematic development of robust machine learning models for a broad range of systems.

\end{abstract}

{\maketitle}

\section{Introduction}

Machine learning has emerged in recent years as a powerful tool
for the description of complex chemical systems~\cite{Behler2016/10.1063/1.4966192,
Bartok2017/10.1126/sciadv.1701816,
Butler2018/10.1038/s41586-018-0337-2,
Deringer2019/10.1002/adma.201902765,
Mueller2020/10.1063/1.5126336}.
A major contribution has been the development of machine learning potentials (MLPs)
--- models that represent potential energy surfaces created by explicit \textit{ab inito} calculations ---
which enables the study of chemical systems
for long timescales and on large length scales, even with chemical reactivity included.
The first method based on
artificial neural networks that is in principle
scalable to arbitrary system sizes was the
high-dimensional neural network
potential (NNP) methodology~\cite{Behler2007/10.1103/PhysRevLett.98.146401,
Behler2017/10.1002/anie.201703114}
combined with atom-centered symmetry
functions to describe atomic environments~\cite{Behler2011/10.1063/1.3553717}.
Over the years, many other distinct methods have been
proposed for this difficult task, based on a range of descriptors and either on
artificial neural networks~\cite{
Ghasemi2015/10.1103/PhysRevB.92.045131,
Khorshidi2016/10.1016/j.cpc.2016.05.010,
Schuett2017/10.1038/ncomms13890,
Artrith2017/10.1103/PhysRevB.96.014112,
Smith2017/10.1039/C6SC05720A,
Zhang2018/10.1103/PhysRevLett.120.143001,
Unke2019/10.1021/acs.jctc.9b00181,
Shao2020/10.1021/acs.jcim.9b00994},
or on kernels~\cite{
Bartok2010/10.1103/PhysRevLett.104.136403,
Rupp2012/10.1103/PhysRevLett.108.058301,
Thompson2015/10.1016/j.jcp.2014.12.018,
Shapeev2015/10.1137/15M1054183,
Li2015/10.1103/PhysRevLett.114.096405,
Chmiela2017/10.1126/sciadv.1603015}.
Since their introduction, NNPs have been successfully applied to
solvents~\cite{Morawietz2016/10.1073/pnas.1602375113,
Morawietz2018/10.1021/acs.jpclett.8b00133,
Cheng2019/10.1073/pnas.1815117116},
solids~\cite{
Artrith2016/10.1016/j.commatsci.2015.11.047,
Artrith2017/10.1103/PhysRevB.96.014112,
Singraber2019/10.1021/acs.jctc.8b01092},
solid-liquid interfaces~\cite{Natarajan2016/10.1039/C6CP05711J},
and reactive processes in solution~\cite{Hellstrom2017/10.1039/C6CP06547C}
or at interfaces~\cite{Quaranta2017/10.1021/acs.jpclett.7b00358,
Hellstrom2019/10.1039/c8sc03033b},
and have therefore repeatedly demonstrated
their reliability for the
understanding of complex molecular systems and materials.
However, a crucial component of any MLP is a robust and representative
training set whose construction can easily become the most challenging
part of the development of such a model, especially for condensed phase systems.

At the same time, it is well known in the
machine learning community that the predictive power
of a machine learning approach can be substantially
improved by combining multiple individual
models~\cite{
Clemen1989/10.1016/0169-2070(89)90012-5,
Hansen1990/10.1109/34.58871,
Krogh1995/10.1.1.37.8876,
Zhao2005/10.1109/icnnb.2005.1614650,
Rokach2010/10.1007/s10462-009-9124-7}.
Instead of a single model,
multiple models are trained independently to form a committee
which offers several benefits.
Averaging over the predictions of an ensemble of committee members usually provides an
improved accuracy of the prediction compared
to the individual members~\cite{
Hansen1990/10.1109/34.58871,
Hashem1995/10.1109/72.377990,
Breiman1996/10.1007/bf00058655,
Hashem1997/10.1016/S0893-6080(96)00098-6}.
In addition, the disagreement of the committee, as measured by
the standard deviation of the predictions of the members, provides access to an estimate of the
generalization error~\cite{
Wolpert1992/10.1016/S0893-6080(05)80023-1,
Krogh1995/10.1.1.37.8876,
Ueda1996/10.1109/icnn.1996.548872}.
Moreover, the committee model can substantially reduce overfitting
issues~\cite{Sollich1996/no-doi}.
Finally, by adding previously unlabeled data with maximal
committee disagreement to the training set, the model can be
systematically improved --- an active learning strategy
known as query by committee (QbC)~\cite{
Seung1992/10.1145/130385.130417,
Krogh1995/10.1.1.37.8876}.

Despite the rise of machine learning in molecular simulations
and materials science, committee models are not considered standard 
tools and have mainly been used in pioneering work.
This includes the well established practice
to use the difference between two NNP models for
the manual improvement of the training
set~\cite{
Artrith2012/10.1103/PhysRevB.85.045439,
Morawietz2016/10.1073/pnas.1602375113,
Chen2020/10.1039/c9ra09935b}
as first described in Ref.~\citenum{Behler2011/10.1039/C1CP21668F},
however without combining the predictions of the two models.
More recently, the use of two machine learning models
has been proposed for the simulation of the infrared spectra of
gas phase molecules either in an ensemble
averaging approach~\cite{Gastegger2017/10.1039/c7sc02267k,Gastegger2020/10.1007/978-3-030-40245-7_12},
or for the error estimation in post-processing~\cite{Raimbault2019/10.1088/1367-2630/ab4509}.
Additionally, the disagreement of NNPs has been shown to be
crucial for the automated fitting of NNPs at
coupled cluster accuracy for protonated water
clusters~\cite{Schran2020/10.1021/acs.jctc.9b00805}.
Ensemble methods were also recently proposed
for uncertainty estimation in chemical machine
learning~\cite{Musil2019/10.1021/acs.jctc.8b00959}.
Besides these examples, QbC strategies have been leveraged for the development of moment tensor potentials~\cite{Podryabinkin2017/10.1016/j.commatsci.2017.08.031}
and more recently for deep potential models~\cite{Zhang2019/10.1103/PhysRevMaterials.3.023804}.
QbC has also been successfully utilized for active learning in chemical space~\cite{Smith2018/10.1063/1.5023802}, which additionally demonstrated
the improvements in accuracy obtained by using the committee average for predictions.
In the realm of Gaussian approximation
potentials~\cite{Bartok2010/10.1103/PhysRevLett.104.136403}
data-driven learning strategies~\cite{Deringer2018/10.1103/PhysRevLett.120.156001}
have recently been shown to be crucial
for the automated development of machine learning potentials.
Finally, complementary strategies such as farthest point
sampling~\cite{Ceriotti2013/10.1021/ct3010563}
have been tested for the construction of uniform data sets~\cite{Musil2018/10.1039/c7sc04665k}.

In this work, we exploit the established
benefits of committee models known in the field of machine learning
to create robust MLPs with controlled generalization errors
in an automated fashion.
Conceptually, our approach is based on
a combination of multiple models of the
well-established NNP formalism~\cite{Behler2007/10.1103/PhysRevLett.98.146401,
Behler2017/10.1002/anie.201703114}.
The resulting committee intentionally shares the same atom-centered symmetry
functions~\cite{Behler2011/10.1063/1.3553717}
as descriptors for the atomic environments,
thus leading to small, often negligible, computational
overhead in production runs.
We show that, compared to the individual NNPs,
this approach results in improved accuracy for predictions and at the same time
gives direct access to the committee disagreement, an estimate of the generalization error.
This disagreement --- being straightforward to
compute during a simulation --- can thus be monitored and
even biased to stabilize the simulation.
It also enables active learning via QbC
techniques which expands the training set,
systematically improving the model.
These benefits allow us to build an adaptive
workflow for the development of committee NNP
models and systematic generation of their training sets.
We finally illustrate the capabilities of the resulting methodology
on the description of water in the condensed phase at
various state points and make the resulting training set
and model parameters available to the community.
This methodology integrates established aspects of committee models with new features such as our adaptive active learning workflow and the biasing of the committee disagreement into a unified framework and will yield robust MLPs for a broad range of systems.

\section{Committee Neural Network Potentials}

Before we introduce the principal ideas underlying committee NNPs (C-NNPs),
we first briefly summarize the original Behler-Parrinello NNP methodology.
To represent an interatomic potential by NNPs,
the atomistic structure is first transformed using
atom-centered symmetry functions~\cite{Behler2011/10.1063/1.3553717}
into translationally and rotationally invariant
descriptors of atomic environments.
These serve as input for atomic neural networks
that output auxiliary components of the total potential energy
which is then obtained
as a sum of contributions from all atoms in the system.
The resulting permutationally invariant structure--energy relation can be
analytically differentiated to obtain forces,
for example to drive molecular dynamics,
and is scalable to essentially arbitrary
system sizes~\cite{Behler2007/10.1103/PhysRevLett.98.146401}.
The whole model is trained by optimizing the parameters (weights and biases)
of the atomic neural networks, one per element, to reproduce the reference
energies and optionally forces
of a training set.
In contrast, the network architecture as well as the particular choice
of symmetry functions are hyperparameters that
need be be specified by the user.
For further details on the original NNP methodology we
refer the reader to Ref.~\citenum{Behler2017/10.1002/anie.201703114}.

In order to extend this methodology to a committee
model, we propose to use multiple NNPs that have been
optimized independently using the same training set.
However, every individual NNP is trained to a slightly
different subset of the full training set, while a small fraction
is intentionally left out in each case.
This strategy, also known as random subsampling
in the machine learning community,
introduces variation between the committee
members as shown, for example, in Ref.~\citenum{Krogh1995/10.1.1.37.8876}
for artificial neural networks.
Together with the intrinsic stochastic nature of
the neural network optimization due to the initialization
of the weights and the optimizer itself, these
different contributing factors provide a
sufficiently diverse committee of NNPs.
Given the predictions of potential energies and atomic forces by the committee of NNPs,
$\{ E_i(q) \}_{i=1}^n$ and
$\{ - \nabla_\alpha E_i(q) \}_{i=1}^n$,
as a function of the positions of all the atoms $q \equiv \{ \vect{q}_\alpha \}_{\alpha=1}^N$,
the C-NNP prediction for any given structure
is obtained as an average,
\begin{equation}
\begin{aligned}
    E(q)
    & =
    \frac{1}{n} \sum_{i=1}^{n} E_i(q),
    \\
    \vect{F}_\alpha(q)
    & =
    \frac{1}{n} \sum_{i=1}^{n} \vect{F}_{i \alpha}(q)
    =
    - \frac{1}{n} \sum_{i=1}^{n} \nabla_\alpha E_i(q),
\end{aligned}
\end{equation}
where $i$-indexed quantities represent predictions of the $n$ individual committee members,
non-indexed ones represent the averaged C-NNP prediction, and $\alpha$ is the atomic index.
As we have a set of predictions for each structure,
we can define the committee disagreement as
the standard deviation of the individual NNPs,
\begin{equation}
\begin{aligned}
    \sigma_E(q)
    & =
    \left[
        \frac{1}{n} \sum_{i=1}^{n} (\Delta E_i)^2
    \right]^{\frac{1}{2}},
    \\
    \sigma_{\vect{F}_{\alpha}}(q)
    & =
    \left[
        \frac{1}{n} \sum_{i=1}^{n} (\nabla_\alpha \Delta E_i)^2
    \right]^{\frac{1}{2}},
\label{eq:force-bias}
\end{aligned}
\end{equation}
where we introduce the notation $\Delta E_i \equiv E - E_i$.
These disagreements can be easily computed and monitored on the fly
during a production run and provide an estimate of the uncertainty
of the C-NNP prediction for a given configuration.
The energy disagreement gives global information,
while the force disagreement is locally resolved
for each atom and can therefore highlight weaknesses
of the prediction for a local environment within a given configuration.
Thus, access to the disagreement enables
direct validation of the predictions of a C-NNP model,
in particular since it is known that the committee disagreement provides
a measure of the generalization error~\cite{
Wolpert1992/10.1016/S0893-6080(05)80023-1,
Krogh1995/10.1.1.37.8876,
Ueda1996/10.1109/icnn.1996.548872}.

At this point, it is clear that due to the correlation
of disagreement and generalization error,
it is beneficial to have small disagreement during
a production run.
This will be the case for a well-trained robust committee model,
but before we obtain one, we can take steps to ensure that disagreement is controlled.
To achieve that, we define a biasing
potential $E^{(\mathrm{b})}$ that acts on the energy disagreement, for example
using a shifted harmonic form
\begin{align}
    E^{(\mathrm{b})}\left[\sigma_E(q)\right]
    &=
    \theta(\sigma_E - \sigma_0)
    \frac{1}{2} k^{(\mathrm{b})}
    \left(
        \sigma_E - \sigma_0
    \right)^2,
\label{eq:total-energy-bias}
\end{align}
where $\theta$ is the Heaviside step function that activates the bias only upon reaching a threshold disagreement $\sigma_0$.
In principle, other functional forms of the biasing potential are possible,
which will be explored in future work.
The above choice makes it particularly easy to compute the associated
biasing forces as
\begin{equation}
\begin{split}
    - \nabla_\alpha E^{(\mathrm{b})}
    &=
    \theta(\sigma_E - \sigma_0) k^{(\mathrm{b})} \frac{\sigma_E - \sigma_0}{\sigma_E}  \\
    & \cdot \frac{1}{n}
    \sum_{i=1}^{n}
    - \Delta E_i
    \nabla_\alpha
    \Delta E_i,
\end{split}
\end{equation}
which can be used to keep the disagreement within reasonable upper
limits in a molecular dynamics run.
Biasing of the committee disagreement therefore
provides a unique way to stabilize a
simulation that employs a committee model.
By shifting the onset of the biasing potential to larger
committee disagreements, the influence on the simulation
can be fine tuned and minimized so that the biasing potential only
acts as a safeguard against rare excursions of very high
disagreement.
Biasing the energy disagreement in this way allows the system
to move freely in parts of configuration space that are well-described by the C-NNP,
while effectively erecting a barrier at the boundary of this region
which prevents the simulation from entering configurations with high generalization errors.
As an alternative which is more local but potentially also more invasive,
separate biases can be introduced on individual atomic force disagreements, as we detail in Appendix A.
We also note in passing that approaches to sample
\textit{intermediate} disagreement, in the spirit of various
enhanced sampling techniques, could provide a new direction
to efficiently generate relevant structures to be included
in training sets of MLPs as part of an active learning procedure.

In the present case, we intentionally decided to
share the same set of symmetry functions for the
representation of atomic environments between the C-NNP
members.
This has the advantage that the evaluation of the
symmetry functions and their derivatives is only performed once for the
whole committee, which is typically the computationally most
demanding step.
Then, only the atomic neural networks are evaluated separately
for each committee member, incurring only small overhead compared to using a single NNP.

\begin{figure}[b]
\centering
\includegraphics{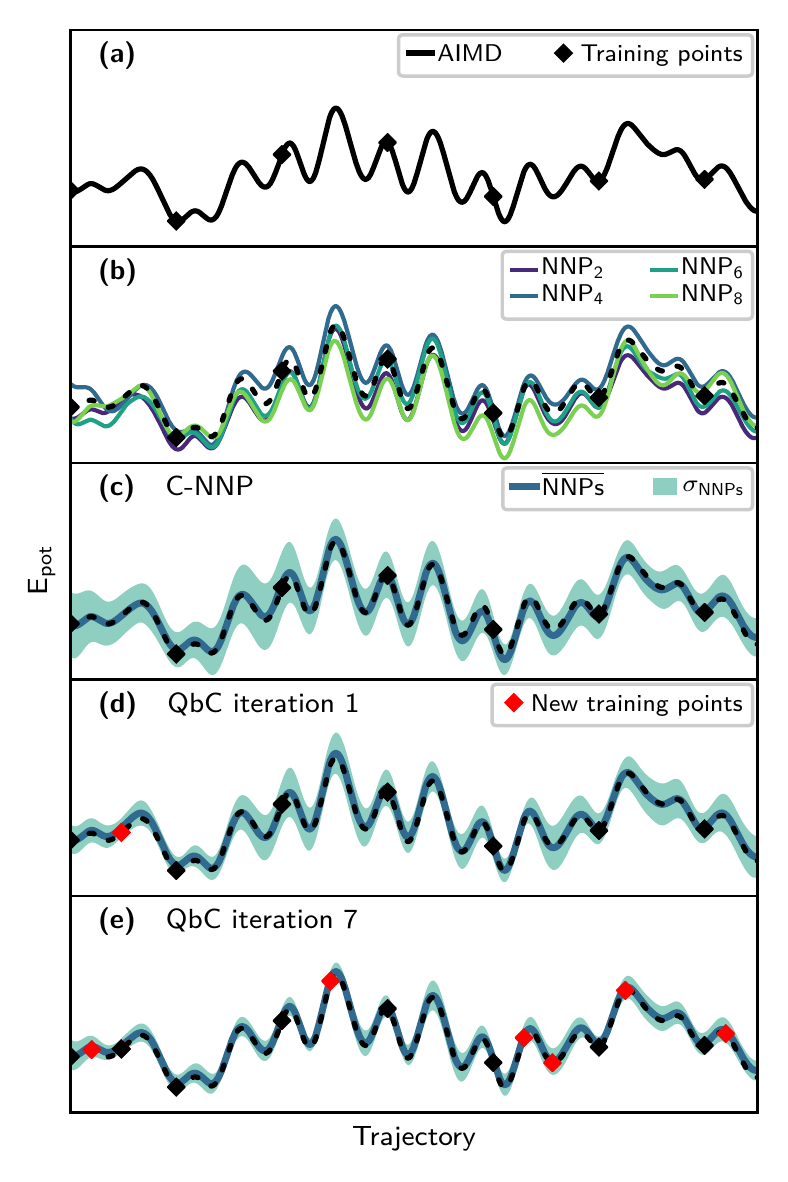}
\caption{\label{fig:committee_model}
Illustration of the committee model compared to
the individual NNP members of the committee and of the QbC procedure
for 64 water molecules in the liquid phase.
(a)
\textit{Ab initio} reference trajectory and first selection
of training points equally spaced along the trajectory.
(b)
Predicted energy along the original trajectory from eight independent
NNP fits to the same training points.
(c)
Predicted energy along the original trajectory of the committee model
composed of the eight NNPs.
(d)
Active improvement of the model via query by committee.
The additional new training point is highlighted in red.
(e)
Performance of the committee model after seven query by
committee iterations.
The committee disagreement has been reduced
and the prediction has improved.
}
\end{figure}

To highlight the benefits of the committee NNP
approach, we illustrate some of its main features
in Fig.~\ref{fig:committee_model}.
A crucial step in the development of any machine learning
potential is the preparation of the training set,
which we address in detail in the next section.
The training set needs to be representative of the planned simulations
and dense enough to generate reliable
interpolation between the training points.
Most preparations of training sets therefore
start from simulations with the
chosen reference method, typically
in the spirit of \textit{ab initio} molecular
dynamics (AIMD)~\cite{Marx2009}.
A first training set can then be generated by
choosing uncorrelated configurations from
such a trajectory as shown in Fig.~\ref{fig:committee_model}A.
Training multiple NNPs with different
initial conditions or to different subsets
of the full training set leads to
varying performance between them,
as highlighted in panel B.
In previous work, the user would then select the
best performing NNP as the model of
choice.
However, if the different models are combined
to a committee NNP, the prediction is substantially
improved, as shown in panel C.
At the same time, the committee disagreement
allows the identification of configurations for which the uncertainty of the model is high,
enabling active learning strategies based on QbC techniques
to iteratively improve the model.
As shown in panels D and E,
adding selected configurations to the training set
substantially reduces the committee's disagreement
while improving its prediction compared to the reference data.
To select new configurations for the training set,
it is possible to use either the global total energy disagreement,
or the local information contained in the atomic force disagreement,
after a suitable reduction over all atoms in the given frame.
Overall, these features allow for a
data-driven approach to developing C-NNP models
as presented in detail in the next section.

The methodology to make use of the benefits
of committee models has been implemented
in the CP2K simulation package~\cite{CP2K}
for Behler-Parrinello NNPs and
will be made available in the next release.
This includes the on-the-fly evaluation of the energy and force
disagreement and the associated biasing of the energy disagreement.
In practice, a committee NNP model can be obtained by performing
individual fits with any NNP training code, for
example with the open-source n2p2 code~\cite{Singraber2019/10.1021/acs.jctc.8b01092}
or the RuNNer code~\cite{Behler/Runner}.
One can therefore see that the proposed concepts are straightforward
to adapt for a broad range of existing MLPs,
while introducing benefits and additional features.

\section{Active Learning Procedure for Committee Neural Network Potentials}

Let us now address a crucial step in the
development of any MLP, the
preparation of the training set.
A machine learning model can only be as good as its
underlying data, which needs to be representative of
the situations encountered when using the final model.
As discussed in the Introduction, the selection of configurations
for the training of machine learning potentials
has recently seen great progress towards data-driven and
automated approaches~\cite{Li2015/10.1103/PhysRevLett.114.096405,
Gastegger2017/10.1039/c7sc02267k,
Podryabinkin2017/10.1016/j.commatsci.2017.08.031,
Smith2018/10.1063/1.5023802,
Deringer2018/10.1103/PhysRevLett.120.156001,
Zhang2019/10.1103/PhysRevMaterials.3.023804,
Chen2020/10.1039/c9ra09935b,
Zhai2020/10.1063/5.0002162,
Lin2020/10.1063/5.0004944,
Schran2020/10.1021/acs.jctc.9b00805}.
In a similar spirit, here we present an adaptive active learning workflow for the construction of robust C-NNPs
for classical and path integral molecular simulations.
The approach developed here builds upon the automated
fitting of NNPs at coupled cluster level of theory
for gas-phase clusters~\cite{Schran2018/10.1063/1.4996819,
Schran2020/10.1021/acs.jctc.9b00805}.

\begin{figure*}
\centering
\includegraphics[width=0.9\linewidth]{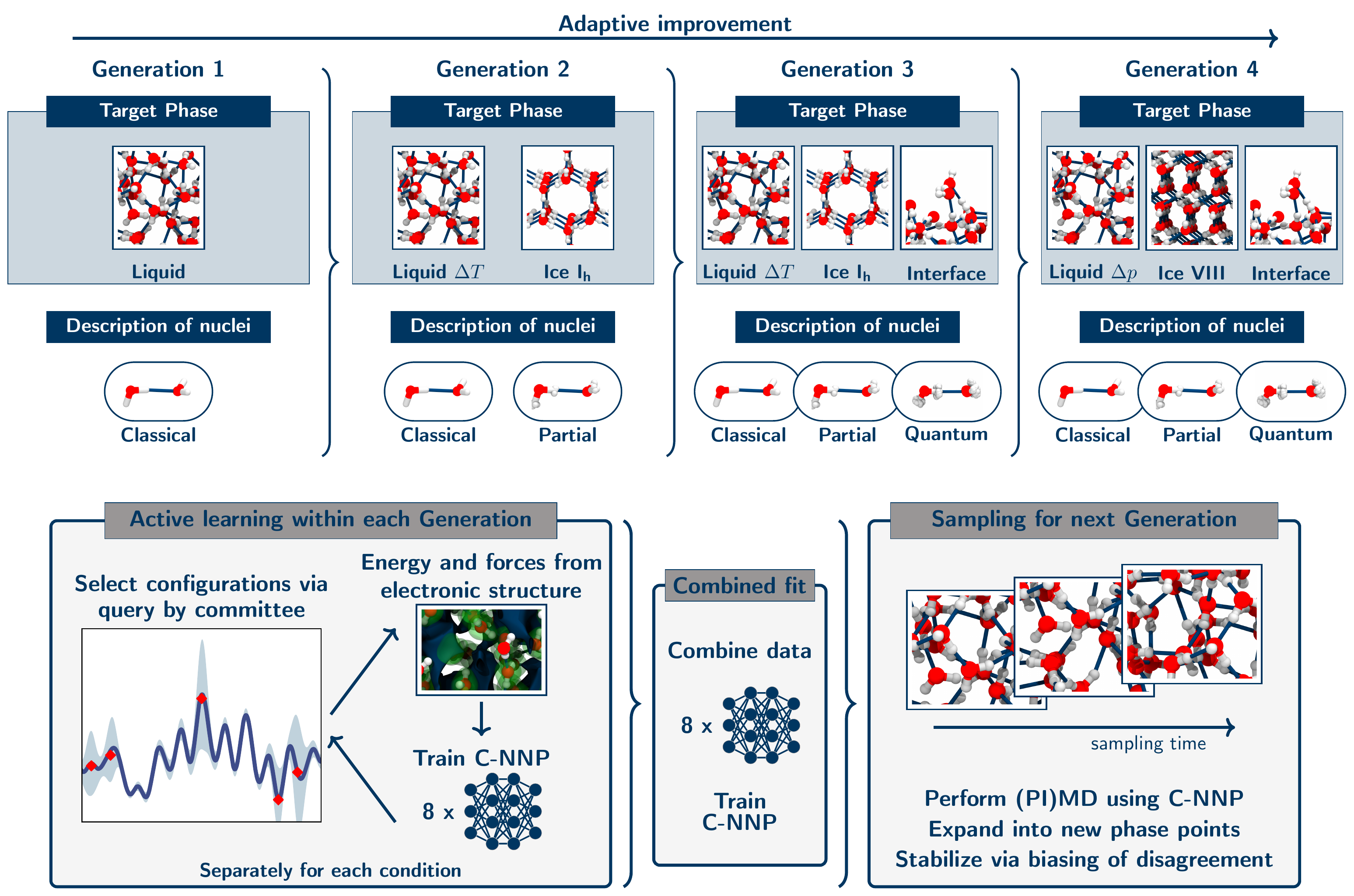}
\caption{\label{fig:systems}
Illustration of the adaptive improvement of the committee NNP over multiple generations.
The top part of the figure summarizes the expansion into new target phases
and the iterative improvement of the description of the nuclei in each generation.
Within each generation the most important points
for an improvement of the model
are actively selected using QbC
based on the highest committee disagreement,
separately for each selected state point (bottom left).
Afterwards, the reference energy and forces, if previously unknown for these structures,
are obtained from explicit electronic structure calculations.
These points are added to the training set and the committee members
are trained to the expanded training set.
QbC iterations are repeated until the committee disagreement converges;
see text for details.
At the end of each generation, all training points are gathered
in order to perform a final extended fit of the committee model
(bottom middle).
The resulting C-NNP can consecutively be applied for exhaustive
(PI)MD sampling at various new state points (bottom right).
These simulations provide the structures for the next generation in
the adaptive improvement.
}
\end{figure*}

We make use of two basic properties of the committee model
to automate the development of C-NNPs.
Firstly, as shown in the previous section,
the committee disagreement
can be used as an estimate of the generalization error
of the model.
By adding configurations to the training points
that feature the highest committee disagreement,
the most important points for an improvement of
the model can be iteratively selected.
This is the main principle behind active learning
via QbC~\cite{Seung1992/10.1145/130385.130417}.
Secondly, the C-NNP is many orders of magnitude
cheaper than the reference electronic structure method
and new configurations can therefore be generated
rapidly using the C-NNP.
These large sets of configurations can then
be efficiently screened using QbC and expensive
reference calculations are only performed
for these selected points.

We organize the active learning workflow into different
generations, each of them comprising multiple QbC cycles and other operations,
as outlined in Fig.~\ref{fig:systems} for the condensed phase of water.
Each generation includes new state points and yields a C-NNP that will be used
to generate new candidate structures for the next generation.
The on-the-fly monitoring and biasing
of the committee disagreement provide invaluable tools to guarantee
the stability of these simulations and the validity of the new configurations.
Only at the beginning of the first generation, the process is seeded
from an AIMD simulation in order to provide an initial set of structures.
If the new conditions are not structurally drastically different
from those in the the previous generation
and we use disagreement biasing to keep molecular dynamics stable,
we can start with a single state point and gradually expand into new regions
without the need to run additional expensive AIMD simulations.
If the final model should be applied
together with a quantum description of the nuclei,
this can also be adaptively included over the
generations by gradually increasing the
quantum character of the nuclei in imaginary time
path integral simulations~\cite{Markland2018/10.1038/s41570-017-0109,
Ceriotti2016/10.1021/acs.chemrev.5b00674}.

Within a generation, QbC
is used to adaptively extended the training set by selecting the most representative configurations
separately for each state point, improving its description,
as schematically shown
in the bottom left panel of Fig.~\ref{fig:systems}.
With this procedure, multiple state points can easily be treated in parallel.
At the very beginning of each QbC cycle, a small number
of random configurations is chosen to train the
first committee,
while in subsequent iterations,
new configurations are selected
based on the highest committee
disagreement.
We chose to use the force disagreement (rather than total energy disagreement)
for this selection, since it is sensitive to the
local environments within a configuration
and insensitive to the global offset of the whole potential energy surface.
Convergence of these individual QbC cycles can be detected by monitoring this disagreement.
If the structures were generated by AIMD simulations,
as is the case at the beginning of the first generation,
the associated reference forces and energies are
already known and the improved C-NNP model can
be trained directly to the growing training set.
In subsequent generations, candidate structures are generated
by molecular dynamics of the previous generation's C-NNP
and explicit electronic structure reference
calculations are only needed for the small
number of actively selected points.
Once all QbC cycles for the selected conditions
in a given generation are converged, the individual
training sets are combined and a final tight optimization
of that generation's resulting C-NNP is performed.

The adaptive improvement of the model and its training set is completed
after several generations, when all desired conditions have been included
and the final C-NNP exhibits the required accuracy
in subsequent production simulations.
In these simulations, the committee disagreement on energy and forces
can be monitored on the fly and compared
to the disagreement known from the active learning process.
If this disagreement stays within the range encountered
for these known conditions included in the training process,
it is expected that the final model reaches the desired accuracy
also in the production simulations.
Thus, utilizing the properties of committee models, the data-driven workflow outlined above
helps automate the development of robust machine learning potentials
and subsequent production simulations with controlled accuracy.

\section{Application of Committee Neural Network Potentials to Water}

\subsection{Development of the Committee Model}
\label{sec:afp_water}

In order to showcase the benefits of the committee NNP
methodology, we develop a C-NNP model for water at
various state points, including also the quantum
nature of the nuclei, following the data-driven workflow described above.
All specific settings used here are listed in the Computational details section, while the training input files, training set and parameters of the
final model are publicly available as specified in the data availability statement.
In the first active learning generation, we seed the procedure with 300 ps of classical AIMD simulation
of liquid water at 300~K obtained at the
hybrid density functional theory (DFT) level~\cite{Marsalek2017/10.1021/acs.jpclett.7b00391}
and perform a single QbC cycle targeting this state point.
This QbC cycle uses a committee of 8 NNPs and is initialized with 20 structures randomly selected from the ensemble.
10 new configurations with the highest disagreement are added in each subsequent iteration.
After the training of the individual members,
the energies and forces of 5000 random structures from
the original trajectory are predicted in order
to compute the committee disagreement.
We intentionally use only a subset of the large pool
of candidate structures in order to make the QbC
iterations computationally more efficient.
For the same reason, the QbC NNPs are optimized
relatively loosely (15 epochs) within each QbC iteration.
In order to select the most relevant configurations
for an improvement of the model, we chose to use
the mean force disagreement of each configuration
to rank the candidate structures.
If a newly selected configuration has already been
included in a previous QbC iteration, it is not added again
to the training set.
Such occasions indicate that the QbC process is reaching
the limits of the provided set of configurations, since
structures are selected more than once.
Given that the DFT energies and forces are already known in
the first generation, the selected points are directly added to the
training set.

\begin{figure}
\centering
\includegraphics{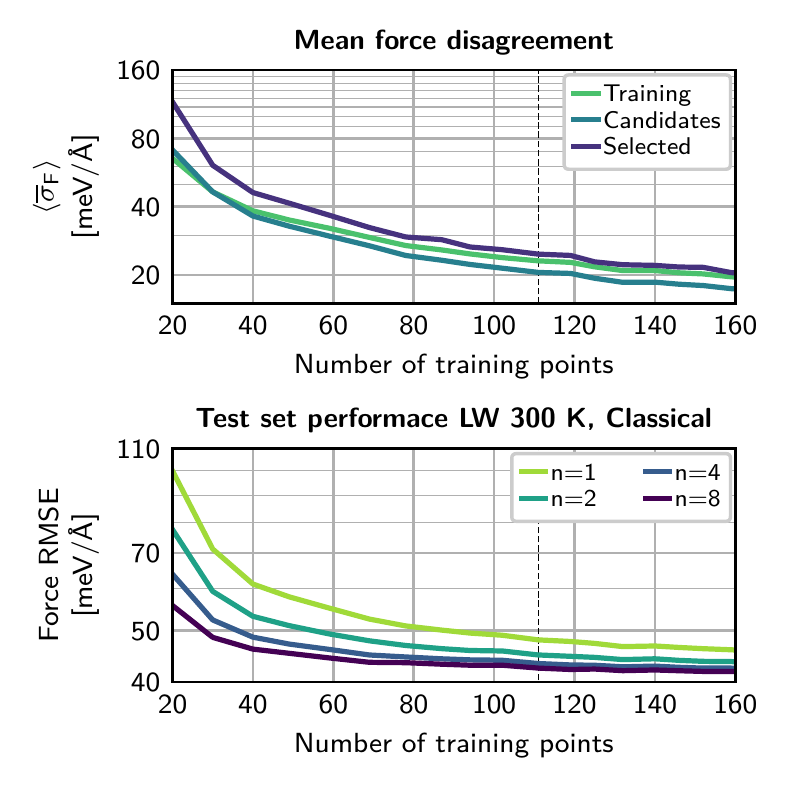}
\caption{\label{fig:afp_performance}
Convergence of the QbC process in the first generation with respect to the number of structures in the training set.
The top panel shows, in logarithmic scale, the mean force committee disagreement
averaged over the given set of structures $\langle\overline{\sigma}_F\rangle$
for the training set (Training),
the large set of potential new candidates (Candidates),
and the actual newly selected configurations (Selected).
The bottom panel shows, in logarithmic scale, the force root mean square error (RMSE) of the C-NNP model
for different committee sizes from one to eight members.
It was evaluated on 500 independently
generated configurations for the target state point of liquid
water at 300~K with classical nuclei;
See Sec.~\ref{sec:comp_det} for details.
The final number of training points (111) used from
the QbC cycle is marked with a vertical dashed line.
}
\end{figure}

Monitoring of the committee disagreement during a QbC cycle
allows the user to easily gauge convergence of the process.
The evolution of the atomic force disagreement
during the first QbC process is shown in the top
panel of Fig.~\ref{fig:afp_performance}
separately for
the structures in the training set,
the newly selected structures,
and the 5000 candidates, from which the next 10 structures for the training set are chosen.
At the beginning of the QbC process, the newly
selected points feature substantially larger disagreement
compared to the large set of candidate structures and
the training set.
As more and more points with highest disagreement are
added to the training set, the disagreement of all
three sets of structures decreases monotonically.
However, the disagreement of the selected points decreases
faster and approaches that of the training set and the
candidate structures, indicating that the newly selected
points are not adding further value
for an improvement of the model anymore.
The disagreement for the training set
and the set of candidates is similar, only slightly higher for the training structures for most of the process,
which shows that the training set picks up the outliers of the ensemble,
but without substantially deteriorating the quality of the model.
The force disagreement for all considered sets of structures
eventually decreases more slowly,
indicating that the active learning
process is well converged after roughly 100 structures
have been added to the training set.

Let us next focus on the actual performance of the C-NNP
model for water at the chosen starting condition.
As mentioned previously, the committee disagreement is
an estimate of the generalization error
and so we should expect the accuracy of the model
to improve over the QbC process
as the disagreement decreases.
In order to validate this expectation for the C-NNP approach,
the evolution of the force root mean square error (RMSE) along the QbC cycle
for an independently generated test set at the
chosen condition is shown in the bottom panel
of Fig.~\ref{fig:afp_performance}.
In addition to the RMSE of the full committee
with eight members, we also include the performance
of all possible committees with four, two, and one member (\textit{i.e.}, individual NNPs) for comparison.
As anticipated from the evolution of the committee disagreement,
the force RMSE of the full eight-member C-NNP starts at roughly
60~meV/\AA{} at the beginning of the QbC process and
converges monotonically to a value of about 40~meV/\AA{}
after roughly 100 points have been added to the training set.
At the same time, the performance of the smaller committees,
and most notably the individual NNPs, is substantially worse,
especially at the beginning, where the individual NNPs
show an RMSE that is twice as large as that of the full C-NNP.
Although the large initial differences decrease as QbC process progresses,
the difference remains clear even when convergence has been reached with roughly 100
training points,
where the committee still outperforms the individual members
and reduces the RMSE from an average of 48~meV\AA{} for
the individual NNPs to 42~meV/\AA{} for the full eight-member C-NNP.
Given the slower convergence with an increasing number of structures,
it is clear that it would take a much larger training set for the individual NNPs
to reach the performance of the C-NNP.
Thus, this analysis highlights the added accuracy of the committee approach,
known from other machine learning applications~\cite{Hansen1990/10.1109/34.58871,
Hashem1995/10.1109/72.377990,
Breiman1996/10.1007/bf00058655,
Hashem1997/10.1016/S0893-6080(96)00098-6}.

Overall, this detailed analysis of the first QbC cycle
shows that only a relatively small number of points
is needed to reach convergence for the starting point
of our active learning procedure.
The 111 structures identified after the first 10 QbC iterations are therefore
used as the final training set of the first generation C-NNP
model.
After stringent re-optimization of the individual NNPs ---
see Sec.~\ref{sec:comp_det} for details --- the
C-NNP model is ready to be used for the generation
of new structures at state points neighboring to the original
ensemble of liquid water at 300~K.
For these simulations, the on-the-fly computation
of the committee disagreement is crucial in order
to judge if the new configurations are physically
meaningful.
In addition, the biasing of the
committee energy disagreement derived above can be used to prevent
the system from entering regions of configuration space
where the model is not well determined by the training set.

\begin{figure}
\centering
\includegraphics{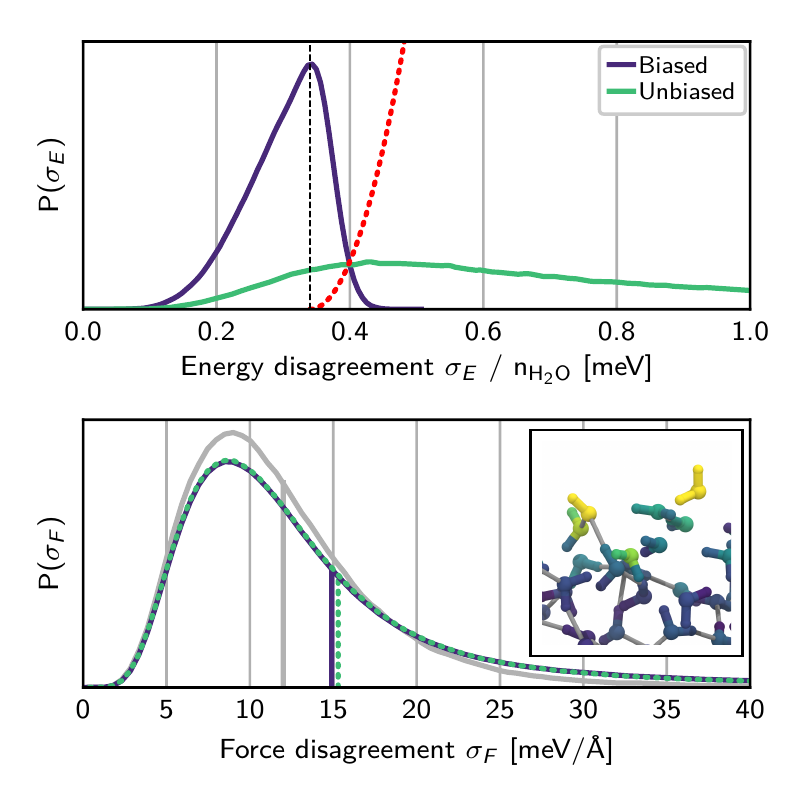}
\caption{\label{fig:com_dis}
Comparison of the distribution of committee disagreement with and without biasing.
The plots show the normalized probability densities of the energy (top panel) and force (bottom panel)
committee disagreement for a water slab
with 216 water molecules at 300~K.
Two simulations were performed for the generation 1 C-NNP model ---
with and without applying a biasing potential acting on the energy disagreement.
The offset of the biasing potential is chosen such that
the bias only acts on configurations with an energy disagreement
per water molecule larger than 0.33~meV per molecule.
The resulting functional form is included in the upper panel
as a red dotted line.
The force disagreement distribution obtained
for the training set of the generation 1 C-NNP model
is shown in gray in the bottom panel and
the averages of the respective distributions are marked
as horizontal lines.
The inset in the bottom panel shows a snapshot
of the air-water interface with atoms color-coded
by their respective force disagreement, where yellow
indicates high and purple indicates low disagreement.
}
\end{figure}

To illustrate the benefits of this feature, we used the C-NNP
model of generation 1 for the simulation of the air-water
interface at 300~K.
In Fig.~\ref{fig:com_dis} we show the resulting probability distributions
of the total potential energy and atomic force disagreement.
The distributions from unbiased simulations feature a very long tail
for the energy disagreement,
which is to be expected from a model that
has not been trained on gas phase clusters or interfaces.
The interface is confirmed as the culprit
by inspecting the spatial distribution of the disagreement,
as shown in the inset in the bottom panel of Fig.~\ref{fig:com_dis},
where individual atoms are colored by their value of atomic force disagreement.
Indeed, the highest values are found for atoms at the interface
whose environments deviate from those in the bulk liquid.
Compared to the force disagreement of the same model for its training set
(gray distribution in the bottom panel of Fig.~\ref{fig:com_dis}),
the distribution from the slab simulation
remains close but exhibits a heavier tail due to the interfacial atoms.
The application of a bias to the energy disagreement suppresses the tail of its distribution
and yields a more compact distribution.
In contrast, it has only a very subtle effect on the distribution of force disagreement,
which highlights the relatively mild influence
of the biasing potential on the local behavior of the system.
Therefore, the energy disagreement biasing can be understood
as a global safeguard that prevents the system from moving into
regions of configuration space with large disagreement,
while keeping local perturbations low.
In light of this analysis, we chose to
use a weak biasing potential for all simulations
used to generate configurations at new conditions,
as detailed in Sec.~\ref{sec:comp_det},
to ensure stability of the simulations without substantial distortion of the structures.

In order to select the target state points for the next
generation in the adaptive improvement of the C-NNP model,
we performed test simulations at a variety of
conditions with the generation 1 model.
After careful analysis of the observed disagreement
for these simulations, we chose
hexagonal ice at 250~K, as well as liquid water
at increased temperatures up to 400~K as the targets
for the second generation of the active learning
process.
Moreover, the quantum character of the nuclei
is targeted by separate PIMD simulations
for the same state points.
To introduce quantum delocalization gradually, we use
underconverged path integral discretization
to stay structurally closer to the
classical ensemble of generation 1.
We employ a separate QbC process to select
new structures for each condition
to ensure that they are optimally covered
by the training set independently of the others.
This has the additional advantage of increased computational efficiency,
as these QbC cycles can easily be run in parallel.
In contrast to the first generation, the
QbC iterations are seeded by choosing 20 random
structures from the training set of the previous
generation.
The DFT reference energy and forces
are unknown for the newly selected configurations
and, thus, are computed during the QbC cycle.
Good convergence of these QbC processes
is reached after only 3--5 iterations and
generation 2 therefore
adds a total of roughly 250 new configurations to the combined
training set from the eight independent QbC cycles.

In the final two generations of the active learning workflow,
additional conditions were targeted.
This includes higher pressure liquid water,
a water slab to represent the air-water interface~\cite{Baer2011/10.1063/1.3633239},
and finally the high pressure ice phase VIII,
as well as the quantum nature of the nuclei,
as summarized in the upper part of Fig.~\ref{fig:systems}.
All details on these target conditions and the
relevant simulations can be found in Sec.~\ref{sec:comp_det}.
During generation 3 and 4, 240 and 205 additional reference
configurations at the various state points were added to the
training set, respectively.
The final result after four generations of our active learning procedure
is a training set of 814 structures and the corresponding tightly optimized C-NNP
able to describe a broad range of conditions with classical or quantum nuclei.
The whole process was originally initialized from a single classical AIMD simulation at 300~K,
with no ab initio path integral molecular dynamics (AIPIMD) required at any point.
This process could
be continued to expand into additional thermodynamic
regions in case they are of interest for specific scientific
questions.
However, we consider the
diversity of the training set sufficient to showcase
the ability of our approach to generate
robust and accurate C-NNP models and their
training sets in a data-driven and automated fashion.

\subsection{Validation of the Committee Model}
\label{sec:validation}

After presenting the details of the active learning
procedure for the development of a C-NNP model
for water at various state points including
quantum nuclei, we analyse the improvement
of the model over the different active learning generations
and validate in particular the quality of
the final generation 4 model.
For that purpose, we explicitly
benchmark static and dynamical thermal properties
against AIMD and AIPIMD simulations
available for a single state point
(liquid water at 300~K)
with and without nuclear quantum effects,
while we compare RMSE values
for all the state points considered.
The RMSE analysis is performed for an independently
created test set which spans the same thermodynamic
state points as targeted during the development of the
model, but has been generated by separate simulations
with the final generation 4 model, as detailed in Sec.~\ref{sec:comp_det}.
This test set comprises a total of 8000 configurations,
split equally between classical and quantum configurations,
for which DFT reference energies and forces have been calculated.
It therefore features one order of magnitude
more configurations than the final training set of generation 4
and enables a comprehensive performance analysis
of the improvement of the model for the various conditions.

\begin{figure}
\centering
\includegraphics{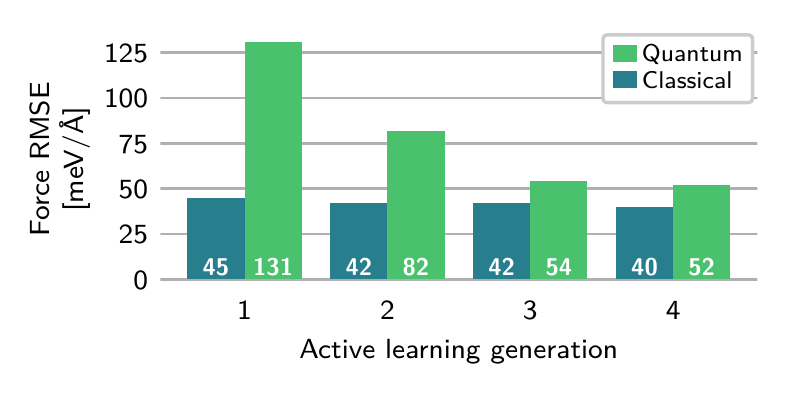}
\caption{\label{fig:gen_imp}
Force root mean square error (RMSE) of the C-NNP
models over the four generations of our active learning workflow.
For each generation, we show the force RMSE relative to the revPBE0-D3 reference
evaluated over an independently generated test set.
The RMSE is averaged over all the state points in each case, separately for classical and quantum structures.
}
\end{figure}

\begin{figure*}[h!t]
\centering
\includegraphics{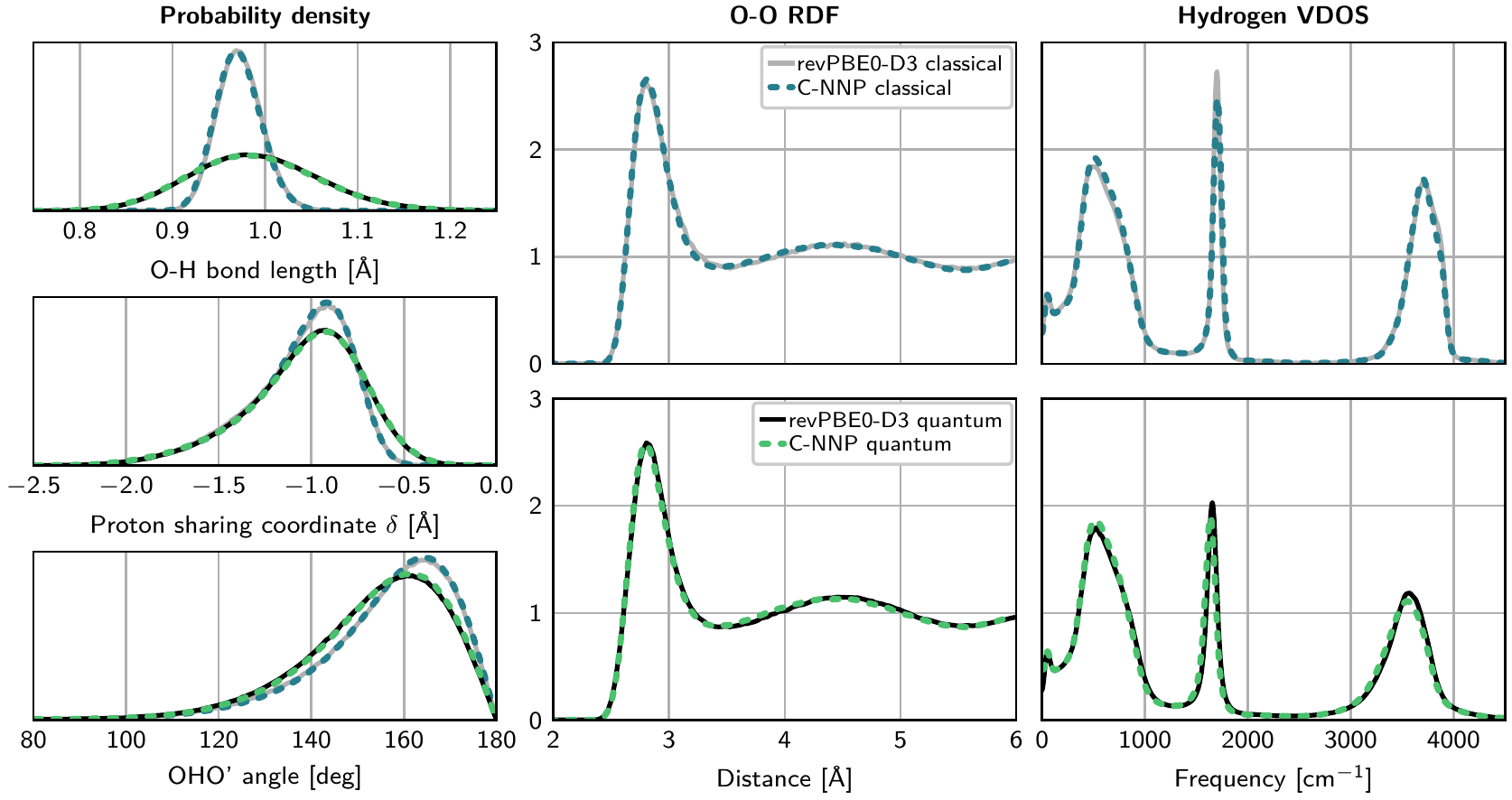}
\caption{\label{fig:all_properties}
Comparison of several local and global static properties
as well as the vibrational density of states
obtained by explicit revPBE0-D3 ab initio simulations
and the final C-NNP generation 4 model.
The three panels on the left show the normalized probability
density for the O-H bond lengths (top), the proton-sharing
coordinate $\delta$ (middle), and the hydrogen bond angle (bottom)
for both a classical and quantum description of the nuclei.
The two panels in the middle display the comparison
of the O-O radial distribution functions, while the
two panels on the right compare the hydrogen atom vibrational
density of states for a classical (top) and a
quantum (bottom) description of the nuclei.
}
\end{figure*}

To summarize the progress over the different generations,
we report in Fig.~\ref{fig:gen_imp} the force RMSE for the independently generated
test set averaged over the various state points but separate for quantum and classical structures.
Classical structures show only minor improvement across the generations,
as they are described already very well by the generation 1 model.
On the other hand, quantum structures,
which were not included at all in the training set of generation 1,
improve substantially with each generation.
The final C-NNP model of generation 4
reproduces the forces in the independent test set
with a RMSE of 40 and 52~meV/\AA{} for classical
and quantum structures, respectively.
This is overall slightly smaller compared to previous
work employing NNPs for water with a classical
description of the nuclei~\cite{
Morawietz2016/10.1073/pnas.1602375113,
Morawietz2018/10.1021/acs.jpclett.8b00133}
and substantially smaller compared to a model
including quantum nuclei~\cite{Cheng2019/10.1073/pnas.1815117116},
while covering diverse regions of the phase diagram and including the quantum
nature of the nuclei with a training set
of just $\sim$800 configurations.
We believe that the surprising robustness of the model with such a small training set
is mainly due to a combination of two factors.
The active learning process selects structures that are the most important for the improvement of the model
while at the same avoiding extreme structures that would distort the fit in regions relevant for simulations.
Furthermore, the fact that the C-NNP is an average of multiple models
lends it stability that is not available to the individual members.

To validate the performance of the final generation 4 C-NNP model
for the calculation of equilibrium properties at a finite temperature,
we compare against reference AIMD and AIPIMD trajectories
of liquid water at 300~K~\cite{Marsalek2017/10.1021/acs.jpclett.7b00391}.
We used the final C-NNP model to obtain two sets of NVT production trajectories,
one with classical nuclei and one with quantum nuclei.
A comparison of both static and dynamical properties
of the system is shown in Fig.~\ref{fig:all_properties}.
Of static properties, we focus first on the local structure
of individual molecules (O-H bond lengths)
and their hydrogen bonds (proton-sharing coordinate $\delta$ and hydrogen bond angle).
The C-NNP model exhibits excellent agreement with the reference AIMD data
for these properties in both the classical and quantum case.
This includes the tail of the quantum distribution of $\delta$ that corresponds
to strong proton sharing, as shown in Fig.~\ref{si_fig:delta}.
We then characterize the intermolecular structure using radial distribution functions (RDFs),
which are, again, captured accurately in both cases.
There is a negligible shift of the second peak of the O-O RDF to shorter distances in the quantum case (Figure~\ref{fig:all_properties}),
while the O-H and H-H RDFs in Figs.~\ref{si_fig:rdfs-oh} and \ref{si_fig:rdfs-hh} show essentially perfect agreement.
Likewise, the vibrational dynamics of the system,
encoded in the vibrational density of states,
is reproduced reliably by the C-NNP model,
as seen for hydrogen atoms in the right column of Fig.~\ref{fig:all_properties}
and in Fig.~\ref{si_fig:vdos-logscale} on a logarithmic scale.
In the classical spectrum, the bending peak at $\sim$1700~cm$^{-1}$ is ever so slightly broadened,
resulting in a small decrease of the peak height, while the rest of the spectrum
is essentially a perfect match, including low-intensity features only visible on logarithmic scale.
In the TRPMD spectrum, the C-NNP model exhibits a very small red shift in the bending and stretching regions
but shows a very good match of the spectrum as a whole.
Note that a contributing factor to the already very small remaining differences in the quantum case
is incomplete convergence of ring polymer contraction in the reference AIPIMD simulations,
as suggested by the comparisons of contraction to 1 and 4 replicas
in the supporting information of Ref.~\citenum{Marsalek2017/10.1021/acs.jpclett.7b00391}.
Overall, we can see that our C-NNP model matches the reference ab initio method accurately
in the description of the structure and vibrational dynamics of liquid water.
This is particularly remarkable in the quantum case,
as no explicit AIPIMD simulations were required in the parametrization of the model.

\begin{figure}[h]
\centering
\includegraphics{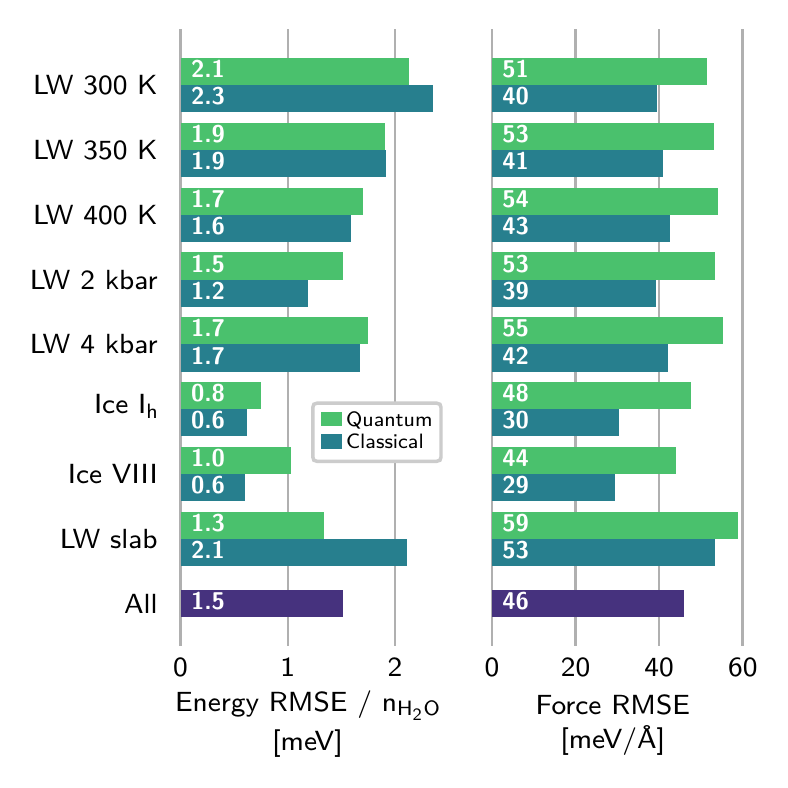}
\caption{\label{fig:test_set}
Root mean square error (RMSE) comparison for the final generation 4 C-NNP model.
The bar chart shows RMSEs of energies and forces
for the C-NNP model with respect
to the revPBE0-D3 reference for an independently
generated test set.
The independent test set consists of 8000 uncorrelated
configurations covering the thermodynamic
conditions targeted during the development of
the C-NNP model and is decomposed into
the individual conditions of
liquid water (LW) at different temperatures and pressures,
hexagonal ice I$_\text{h}$ and ice VIII, as well as
the air-water interface (LW slab),
separately for a classical and quantum description of the nuclei.
See Sec.~\ref{sec:comp_det} for details on the creation of the independent test set.
}
\end{figure}

In the next step, we again widen the analysis of the
performance of the developed C-NNP model for water.
As just demonstrated, it is certainly possible to explicitly validate
various properties with respect to AIMD, or even
AIPIMD simulations at selected conditions.
However, this type of analysis quickly gets out
of scope for all targeted state points that we
considered in the present case.
In addition, the whole purpose of the development
of MLPs is usually to replace expensive ab initio
sampling.
We therefore come back to a detailed analysis of the
RMSEs for energies and forces for the different
conditions accessible in our independent test set.
As seen in Fig.~\ref{fig:test_set}, the final
model scores overall quite well for all considered
conditions, usually with slightly larger RMSE
values for quantum structures.
The water slab performs the worst in
this analysis, indicating the more complex
nature of these configurations due to
the presence of an interface, but still features
RMSE values similar or lower than reported
for previous work on water~\cite{
Morawietz2016/10.1073/pnas.1602375113,
Morawietz2018/10.1021/acs.jpclett.8b00133,
Cheng2019/10.1073/pnas.1815117116}.
At the same time, the two phases of ice
are reproduced best, as expected from the simpler
nature of these systems based on the
arrangement of the molecules on a lattice.
Although nuclear quantum effects lead to
substantially larger and broader potential energy
distributions, the performance of the
final model is convincingly good across the board.
Comparing the score of liquid water for
classical and quantum nuclei at 300~K
to the other conditions makes us confident that
also the associated properties are reproduced
with similarly convincing agreement to
explicit ab initio sampling.

Having seen the performance of the final model,
it is worthwhile to step back and see how it emerges from the active learning process.
To do that, we repeat the same validation procedure as above for the model from generation 1,
\textit{i.e.}, a C-NNP trained on 111 structures selected from classical AIMD trajectories
of liquid water at 300~K using QbC.
The results of comparison to explicit AIMD and AIPIMD for equilibrium properties
shown in Figs.~\ref{si_fig:all_properties} and Fig.~\ref{si_fig:delta_gen1}
clearly demonstrate that even this model does very well in simulations.
For classical nuclei, it is, in fact, as good as the more extensively trained final model.
Remarkably, it performs well even in the quantum case,
despite the fact that no path integral structures at all were used in its training.
The only noticeable deviations are the red shift
in the infra red region of the vibrational spectrum
and the increased proton sharing,
both only slightly larger than for the final model.
We have already seen in Fig.~\ref{fig:afp_performance} how the RMSE against revPBE0-D3 energies and forces
evaluated on the independent test set for liquid water at 300~K
converges during the QbC process that generates the training set for this model.
When we look at the RMSEs also for other conditions in Fig.~\ref{si_fig:test_set},
it is clear that the generation 1 model performs worse than the final one,
especially for quantum structures, which are entirely absent from the training set.
However, forces for classical structures are at the level of the generation 4 C-NNP,
with the exception of ice VIII and the air-water interface.
These states are not only absent from the small training set of the generation 1 model,
but also structurally substantially different from liquid water and so
it should be expected that they are represented less accurately.
This combination of performance characteristics offers an opportunity
to gauge the meaning, or possibly the limitations, of RMSEs when it comes to molecular dynamics.
While the force RMSEs are roughly three times as high for quantum structures as they are for classical ones for the generation 1 model,
these forces are sufficient to generate path integral trajectories that match the reference almost as well as in the classical case.
In terms of computational cost, the 10 iterations of a single QbC process selecting structures from a pre-existing AIMD trajectory
took 6.7 hours on a single AMD EPYC computational node with 32 cores.
Given that this C-NNP is rather easy and computationally cheap to create,
this seems to offer an efficient way to extend existing ab initio simulations
using an MLP from which we do not expect broad generality.

\section{Conclusions}

In summary, we have shown how committee models
can be exploited in the context of machine learning potentials,
which we demonstrated specifically using
Behler-Parrinello neural network potentials,
obtaining a compact training set and a robust C-NNP model with a range of advantages.
While the committee prediction outperforms its individual member NNPs,
the committee disagreement provides a way to monitor and control the accuracy of the model relative to its parent ab initio method and its training set.
Using a bias of the committee disagreement that we introduced,
C-NNP simulations
can be stabilized by effectively erecting a barrier at the boundary
of regions of configuration space described well by the training set,
thus preventing such simulations from entering regions of high uncertainty and
high generalization errors.
The computational overhead of our approach is low compared to using a single NNP,
as we intentionally share descriptors of atomic environments
between the committee members,
which means only the atomic neural networks need to be evaluated multiple times.

Making use of the committee disagreement and repeated query by committee processes,
we designed an active learning procedure capable of generating training sets
in a largely automated and data-driven fashion
while also keeping the number of required reference ab initio calculations to a minimum.
To demonstrate the benefits of this methodology,
we developed a C-NNP machine learning potential ---
made freely available ---
for water under a variety of conditions
with both a classical and a quantum description of the nuclei.
Even though its training set comprises only $\sim$800 configurations,
the final model shows excellent agreement
with the DFT reference method
in comparisons of energies and forces as well as
in comparisons of properties calculated in classical and quantum molecular dynamics simulations.

The promising results presented in this work
make us confident that the methodology introduced here
can be used to systematically develop robust and general C-NNP models
applicable across broad regions of the phase diagram and under a variety of conditions
for systems of increasing complexity.
Given the remarkable performance of the first
generation C-NNP model trained on a small number of samples from a single
AIMD simulation, we see great potential in the development of simpler C-NNP models
for the direct extension of ab initio simulations for a given state point.
Thanks to the much smaller size of the resulting training set compared to previous work,
it will be possible to use more advanced reference ab initio methods in the condensed phase.
Finally, we expect that the described concepts
can be applied directly to other MLPs based on artificial neural networks
and can be also transferred to kernel-based MLPs after the introduction
of stochastic elements into their training process.
We hope that thanks to the low additional complexity and required effort,
committee-based models can become a routine component of the development of machine learning potentials.

\section{Computational Details\label{sec:comp_det}}

We implemented the active learning workflow in Python,
interleaving data manipulation and execution
of various simulation packages to perform specific tasks,
such as the optimization of individual NNPs,
the evaluation of DFT reference energies and
forces for selected configurations, and the
sampling of new configurations.
With this code, the workflow for the development of a C-NNP
model for water described in Sec.~\ref{sec:afp_water}
was carried out over four generations.

All NNP optimizations were performed with the
open-source n2p2 code~\cite{Singraber2019/10.1021/acs.jctc.8b01092}
and, unless stated otherwise, all of the optimization parameters have
been chosen according to the detailed benchmarking
of this code for water~\cite{Singraber2019/10.1021/acs.jctc.8b01092}.
We decided to use an established set of symmetry functions
which has been shown to be able to reproduce the properties
of water over a large range of conditions~\cite{Morawietz2016/10.1073/pnas.1602375113}.
The values of each symmetry function were centered
around the respective average value of the training set and normalized to
values between zero and one.
These atomic environment vectors serve as the input for the atomic NNs
consisting of two hidden layers of 20 neurons each with
hyperbolic tangent activation functions for these two hidden layers
and a linear activation function for the output neuron.
The eight-member C-NNP models were constructed by random subsampling
of the full set of reference data
for each committee member, where 10\% of the points were
randomly left out in each case.
After different random initialization for each committee member,
the weights and biases of the NNs were optimized using the
parallel multistream version~\cite{Singraber2019/10.1021/acs.jctc.8b01092}
of the adaptive global extended Kalman filter~\cite{Shah1992/10.1016/S0893-6080(05)80139-X,
Blank1994/10.1002/cem.1180080605}
as implemented in n2p2.
C-NNPs used for QbC were optimized for 15 epochs with 8 streams,
while the final C-NNPs for each generation, to be used for simulations,
were optimized for 100 epochs with 32 streams.

The DFT reference calculations were all performed
with the CP2K simulation package~\cite{CP2K,Hutter2014/10.1002/wcms.1159}
and its Quickstep module~\cite{VandeVondele2005/10.1016/j.cpc.2004.12.014}
using the exact same electronic structure setup
for the revPBE0-D3 functional~\cite{Perdew1996/10.1103/PhysRevLett.77.3865,Zhang1998/10.1103/PhysRevLett.80.890,Adamo1999/10.1063/1.478522,Grimme2010/10.1063/1.3382344} employed in
Ref.~\citenum{Marsalek2017/10.1021/acs.jpclett.7b00391}.
As shown therein, this choice of functional provides
reliable properties of water, especially with quantum nuclei,
and is therefore the ideal
choice for the development of our C-NNP model.

The simulations using the C-NNP models
used to generate candidate structures for the next active learning generation
were all performed with a modified version
of the CP2K simulation package~\cite{CP2K,Hutter2014/10.1002/wcms.1159}.
All classical simulations were propagated for 0.5~ns
with a time step of 0.5~fs and a global CSVR thermostat~\cite{Bussi2007/10.1063/1.2408420}
with a 30~fs time constant to sample the NVT ensemble.
From these simulations, every 20th configuration
was saved to ensure uncorrelated statistics,
resulting in 50000 configurations for all
classical ensembles.
All PIMD simulations were propagated for 0.25~ns
with a time step of 0.25~fs and a PILE thermostat~\cite{Ceriotti2010/10.1063/1.3489925}
to sample the quantum canonical ensemble.
Partially converged path integral simulations were
realized with 4 replicas to discretize the path integral, which
corresponds approximately to the midpoint between
classical simulations and
converged path integral simulations,
as shown for example in Ref.~\citenum{Schran2018/10.1021/acs.jctc.8b00705}
for the prototypical hydrogen bond in the Zundel cation.
Full PIMD simulations were performed with
16 replicas for all simulations except at
250~K, where 32 replicas were used.
In all quantum cases, every 40th configuration from
4 path integral replica trajectories was used to generate
the set of candidate structures for the QbC cycles, resulting in a total
in 100000 configurations for every quantum ensemble.
After careful testing of the influence of the
energy committee disagreement biasing derived above,
mild biasing with a harmonic constant $k^{(\mathrm{b})}=0.95/\text{eV}$
per atom and a shift of $\sigma_0 = 0.1~\text{meV}$ per atom
was used for all simulations during the active learning process.
The C-NNP model of generation 1 was used for
the simulation of liquid water along an isochor
in a cubic box of size 12.42~\AA{} with 64 water molecules
at three temperatures of 300, 350, and 400~K.
The proton disordered phase ice I$_\text{h}$
was simulated at 250~K in supercells with periodic
boundary conditions including 96 water molecules,
where the lattice size and initial conditions
were chosen according to Ref.~\citenum{Schran2019/10.1039/C9CP04795F}.
These 4 state points were targeted both with
a classical and partially converged quantum description
of the nuclei for the generation 2 C-NNP model.
The resulting C-NNP model of generation 2 was used to
perform converged PIMD simulations along the
same isochor of liquid water as well as
for ice I$_\text{h}$.
Besides these 4 simulations, a water slab
was simulated with 216 molecules in a 15$\times$50$\times$15~\AA{}
periodic box starting from an initial condition
from Ref.~\citenum{Vacha2011/10.1021/jz2014852} both with classical
and partially converged quantum nuclei.
Finally, the generation 3 C-NNP model was
employed to simulate liquid water along an isotherm
for two pressures of 2 and 4~kbar
resulting in cubic boxes with sizes of 12.13 and 11.93~\AA{}
for 64 water molecules.
In addition, ice VIII was simulated at 250~K
with 64 molecules in a 9.70$\times$9.70$\times$14.11~\AA{}
periodic box.
These 6 simulations were both performed
for classical and converged quantum nuclei.
Finally, the water slab was simulated
as in the previous generation, but now with
converged quantum nuclei.

In order to generate an independent test set to
validate the performance of the C-NNP models
at the individual generations and the various state points,
we used the final C-NNP
model of generation 4 to perform independent
simulations at all previously targeted conditions
and state points.
All classical and quantum simulations were propagated for
50~ps and 25~ps, respectively, with otherwise
identical settings as for the above mentioned simulations.
The path integral has been fully converged
for all these simulations by using 128 replicas
for all simulations at and above 300~K and
256 replicas for the ice phases at 250~K.
The biasing of the energy committee disagreement
has been switched off for these production runs.
From these 8 classical and 8 quantum
ensembles in total, 500 uncorrelated structures
were extracted in each case and the reference revPBE0-D3
energies and forces were evaluated.
This set of 8000 reference calculations, spanning
very different conditions both for classical
and quantum nuclei, was consequently used to
validate the performance of the C-NNP models.

Finally, the performance of the generation 1 and 4 C-NNP models
for static and dynamical properties was benchmarked
against previously published AIMD and AIPIMD
results for liquid water~\cite{Marsalek2017/10.1021/acs.jpclett.7b00391}.
For that purpose, 64 water molecules in a cubic box of
size 12.42~\AA{} were simulated at 300~K in the NVT ensemble
to match exactly the setup of the reference ab initio simulations.
We performed classical simulations with both C-NNP models,
using a time step of 0.5~fs and
a global CSVR thermostat~\cite{Bussi2007/10.1063/1.2408420} with a time constant of 1~ps.
Results including nuclear quantum effects
were obtained from TRPMD simulations~\cite{Ceriotti2010/10.1063/1.3489925,Rossi2014/10.1063/1.4883861}
with 32 path integral replicas using a 0.25~fs time step.
We accumulated a total simulation length of 2~ns
in each case for both models.
Again, no biasing of the committee disagreement
was used for these production simulations.

\section*{Supplementary material}

Additional benchmarking results for the final generation 4 C-NNP
model and detailed analysis of the performance of the C-NNP model
obtained in generation 1 can be found in the Supplementary Material.

\begin{acknowledgments}

This work was supported by the Primus16/SCI/27/247019 grant from Charles University.
This work was partially supported by OP RDE project No. CZ.02.2.69/0.0/0.0/18\_070/0010462, International mobility of researchers at Charles University (MSCA-IF II).
This work was supported by the project SVV 260586 of Charles University.
KB acknowledges funding from the IMPRS for Many Particle Systems in Structured Environments.
Part of this work was supported by the \textit{Alexander von Humboldt--Stiftung} through a research grant awarded to CS.
\end{acknowledgments}

\section*{Data availability}

The data that support the findings of this study are openly
available in \texttt{paper-c-nnp} at \url{http://doi.org/10.5281/zenodo.4004590}.


{
\appendix

\section{Biasing of Force Disagreement\label{sec:bias_force}}

An alternative approach to biasing due to committee disagreement
that provides sensitivity to local structural changes can be based on
the force disagreement introduced in Eq.~\ref{eq:force-bias}.
Rather than bias the total energy disagreement,
one can bias the disagreement of each atomic force vector separately.
Atom $\alpha$ that enters a region with a force disagreement larger than $\sigma_0$
brings a contribution to its biasing energy into the system
given again by a shifted harmonic form as
\begin{equation}
    E_{\alpha}^{\mathrm{(b)}}
    =
    \theta(\sigma_{\mathbf{F}_\alpha} - \sigma_0)
    \frac{1}{2} k^{\mathrm{(b)}}
    \left(
        \sigma_{\mathbf{F}_\alpha} - \sigma_0
    \right)^2.
\end{equation}
This affects all the other atoms in the system, say $\beta$,
and the corresponding biasing force exerted on them is
\begin{equation}
\begin{split}
     \mathbf{F}_{\beta\alpha}^{\mathrm{(b)}}
     &= - \nabla_\beta E_{\alpha}^{\mathrm{(b)}} \\
     &= - \theta(\sigma_{\mathbf{F}_\alpha} - \sigma_0) k^{\mathrm{(b)}} (\sigma_{\mathbf{F}_\alpha} - \sigma_0) \nabla_\beta \sigma_{\mathbf{F}_\alpha}.
\end{split}
\end{equation}
At this point, we have to express $\nabla_\beta \sigma_{\mathbf{F}_\alpha}$
in order to obtain a useful expression for the biasing force.
Differentiating through the definition given in Eq.~\ref{eq:force-bias}, we get
\begin{equation}
\begin{split}
    \nabla_\beta \sigma_{\mathbf{F}_\alpha}
    &= \nabla_\beta \left[
        \frac{1}{n} \sum_{i=1}^{n} (\nabla_\alpha \Delta E_i)^2
    \right]^{\frac{1}{2}} \\
    &= \frac{1}{n} \frac{1}{2 \sigma_{\mathbf{F}_\alpha}} \sum_{i=1}^{n} \nabla_\beta
    \left[
        (\nabla_\alpha \Delta E_i) \cdot (\nabla_\alpha \Delta E_i)
    \right]
    \\
    &=
    \frac{1}{n} \frac{1}{\sigma_{\mathbf{F}_\alpha}} \sum_{i=1}^{n}
        (\nabla_\alpha \Delta E_i) \cdot (\nabla_\beta \nabla_\alpha \Delta E_i)
\end{split}
\end{equation}
as the desired result.
A second derivative matrix operator $\nabla_\beta \nabla_\alpha$ appears,
which generally requires an evaluation of analytical or numerical second derivatives for all $3N$ degrees of freedom
and thus its applicability for practical calculations in this raw form is limited.
However, we note that in this case the second derivative is projected on the direction of the force $\mathbf{F}_{i \alpha}$.
This suggests that finite difference methods that limit the number of numerical derivative evaluations
to only the one in the desired direction to evaluate the directional derivative
could result in a practical computation scheme~\cite{Kapil2016/10.1063/1.4971438,Buchowiecki2013/10.1016/j.cplett.2013.09.070}.
Still, in addition to having to deal with second derivatives,
this biasing has to be evaluated separately for each atom $\alpha$
and would therefore incur a substantial computational cost.
For this reason, we have not implemented it
and used only the biasing of the total energy disagreement given by Eq.~\ref{eq:total-energy-bias},
which worked sufficiently well in practical simulations.

}

\section*{References}

%


\clearpage

\setcounter{section}{0}
\setcounter{equation}{0}
\setcounter{figure}{0}
\setcounter{table}{0}
\setcounter{page}{1}

\renewcommand{\thesection}{S\arabic{section}}
\renewcommand{\theequation}{S\arabic{equation}}
\renewcommand{\thefigure}{S\arabic{figure}}
\renewcommand{\thepage}{S\arabic{page}}
\renewcommand{\citenumfont}[1]{S#1}
\renewcommand{\bibnumfmt}[1]{$^{\rm{S#1}}$}

\title{Supplementary material for: \mytitle}
{\maketitle}

\onecolumngrid

\section{Additional validation of the committee model}

In this section we provide additional benchmarking results for the final generation 4 C-NNP model
that complement what is discussed in Sec.~\ref{sec:validation} of the main text.
Overall, all comparisons of the model against
AIMD and AIPIMD results of liquid water at
300~K highlight the excellent agreement
between the C-NNP model and the reference
ab initio trajectories.

\begin{figure}[h!]
\centering
\includegraphics{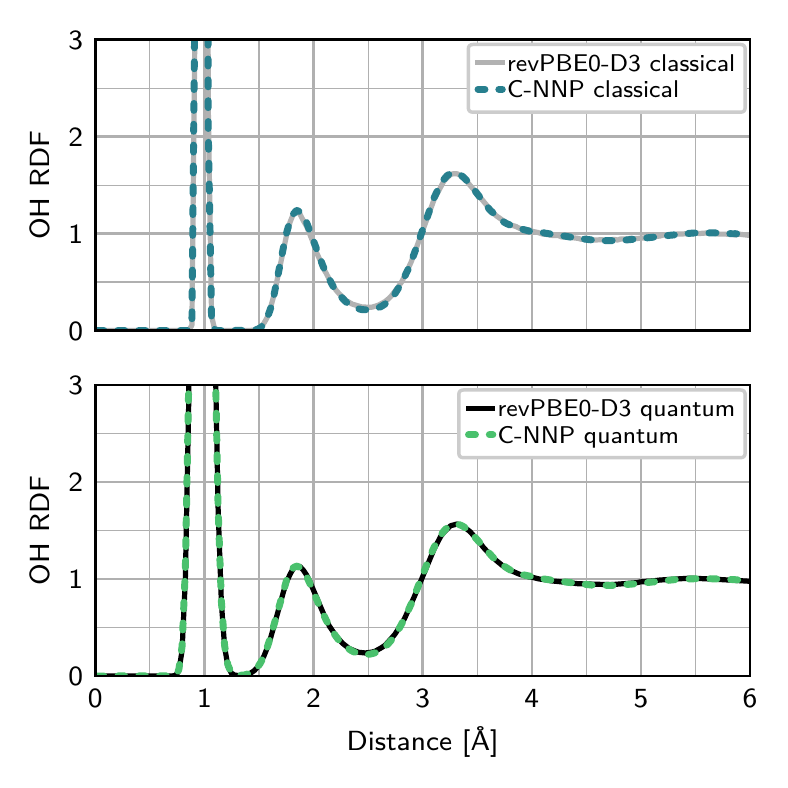}
\caption{\label{si_fig:rdfs-oh}
O-H RDFs complementary to the O-O RDFs shown in Fig.~\ref{fig:all_properties} for the generation 4 C-NNP model for both classical and quantum simulations of liquid water at 300~K.
}
\end{figure}

\begin{figure}[h!]
\centering
\includegraphics{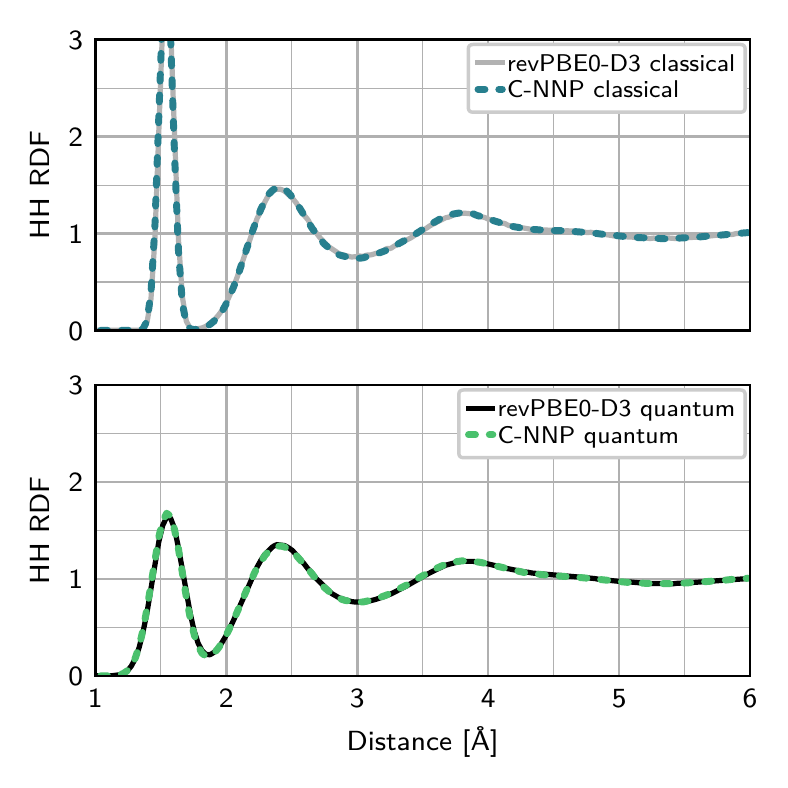}
\caption{\label{si_fig:rdfs-hh}
H-H RDFs complementary to the O-O RDFs shown in Fig.~\ref{fig:all_properties} for the generation 4 C-NNP model for both classical and quantum simulations of liquid water at 300~K.
}
\end{figure}

Figures~\ref{si_fig:rdfs-oh} and~\ref{si_fig:rdfs-hh} show RDFs for O-H and H-H pairs.
The generation 4 C-NNP model yields results that overlap perfectly with the reference data in both the classical and the quantum case.

\begin{figure}[h!]
\centering
\includegraphics{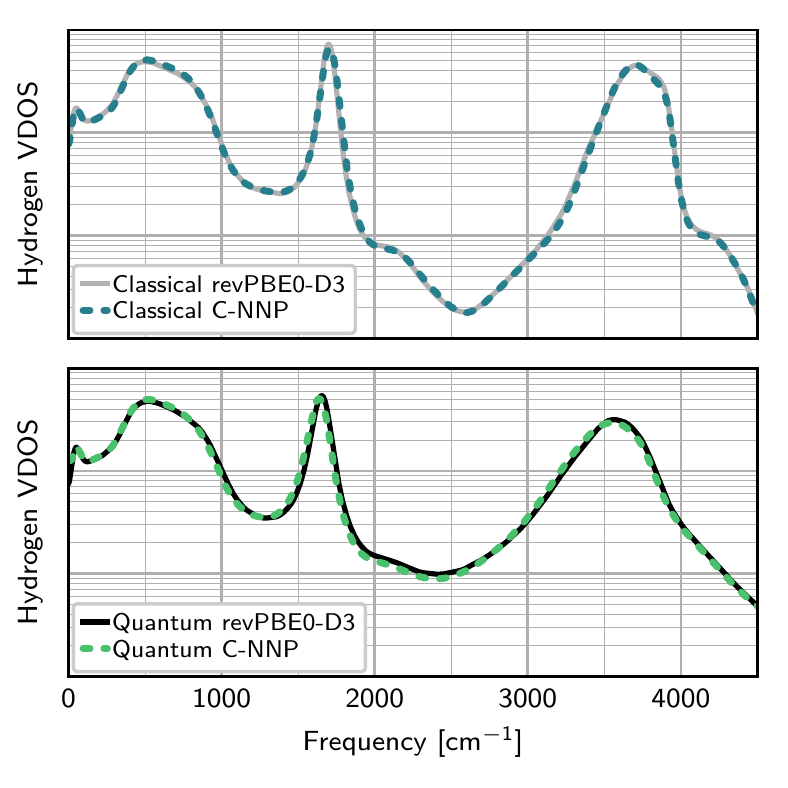}
\caption{\label{si_fig:vdos-logscale}
Hydrogen VDOS in logarithmic scale obtained from the generation 4 C-NNP model simulations.
}
\end{figure}

Fig.~\ref{si_fig:vdos-logscale} shows the hydrogen VDOS discussed in the main text (Fig.~\ref{fig:all_properties}) in logarithmic scale to reveal that also low-intensity features of the spectrum, which are hardly distinguishable in linear scale, are reproduced accurately by the generation 4 C-NNP model.

\begin{figure}[h!]
\centering
\includegraphics{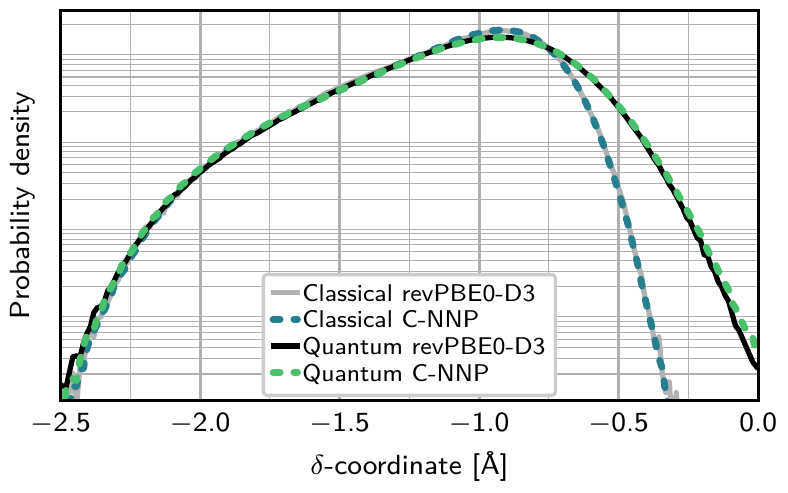}
\caption{\label{si_fig:delta}
Distributions of the proton-sharing coordinate $\delta$ for the generation 4 C-NNP model as obtained with classical and quantum nuclei for liquid water at 300~K, shown in logarithmic scale.
}
\end{figure}

Fig.~\ref{si_fig:delta} shows the distribution of the proton-sharing coordinate $\delta$ in logarithmic scale for the generation 4 C-NNP model.
This highlights the performance of the C-NNP model in the tails of the $\delta$ distribution, namely in the region close to zero that corresponds to strong proton sharing between a pair of water molecules.
Again, the model essentially exhibits a perfect match in the case of classical nuclei all the way along the tails.
The quantum distribution is captured accurately as well, suggesting that the final C-NNP model provides not only a robust molecular structure but also a reliable description of proton sharing phenomena with quantum nuclei.

\twocolumngrid

\FloatBarrier

\onecolumngrid

\clearpage

\section{Validation of Generation 1 Committee Model}

In this section, we repeat the same analysis of static
and dynamical properties for the generation 1 C-NNP
model as presented in detail in the main text for
the final generation 4 C-NNP model.
For that purpose, we again compare the results obtained
from simulations with the generation 1 C-NNP model
for the calculation of equilibrium properties at a finite temperature
against reference AIMD and AIPIMD trajectories of liquid water at 300 K.

\begin{figure*}
\centering
\includegraphics{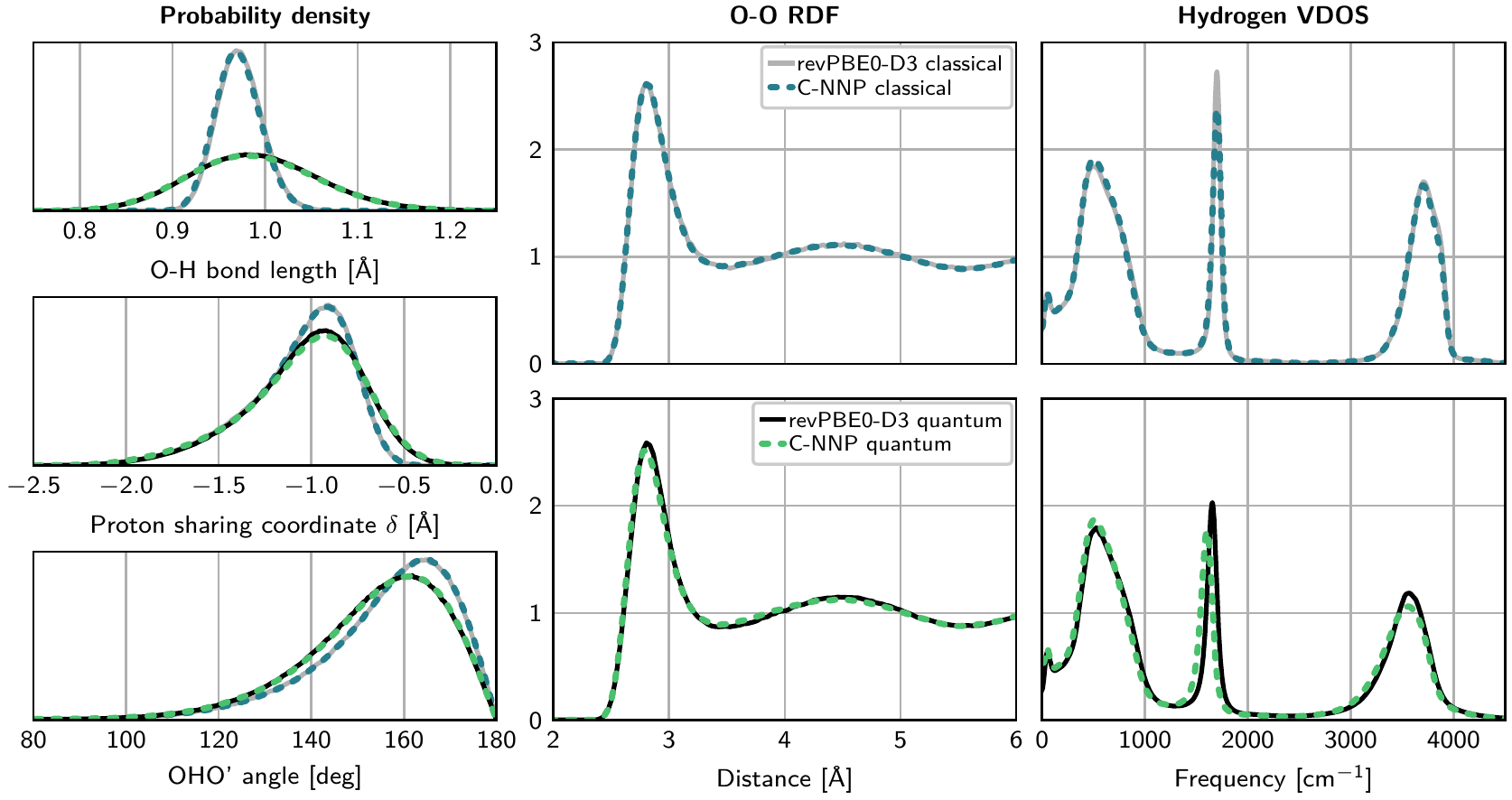}
\caption{\label{si_fig:all_properties}
Comparison of several local and global static properties
as well as the vibrational density of states
obtained by explicit revPBE0-D3 ab initio simulations
and the final C-NNP generation 1 model.
The three panels on the left show the normalized probability
density for the O-H bond lengths (top), the proton-sharing
coordinate $\delta$ (middle), and the hydrogen bond angle (bottom)
for both a classical and quantum description of the nuclei.
The two panels in the middle display the comparison
of the O-O radial distribution functions, while the
two panels on the right compare the hydrogen atom vibrational
density of states for a classical (top) and a
quantum (bottom) description of the nuclei.
}
\end{figure*}

Plots capturing the performance of generation 1 C-NNP model
for the various properties discussed in Sec.~\ref{sec:validation} of the main text
are shown in Fig.~\ref{si_fig:all_properties} together with the original AI(PI)MD reference data.
In comparison with the data shown in Fig.~\ref{fig:all_properties} for generation 4,
it is clear that the generation 1 model already reaches the desired level of accuracy
for the classical static and dynamic properties of liquid water at 300~K.
At the quantum level, the C-NNP predicted properties exhibit minor but visible deviations
from the corresponding AIPIMD data, namely in the O-O RDF and VDOS,
where the O-H stretching and bending peaks are broadened and red-shifted to different degrees.
However, the overall match can still be considered remarkably accurate,
understanding that no quantum structures were used in the training of the generation 1 model.

\begin{figure}[h]
\centering
\includegraphics{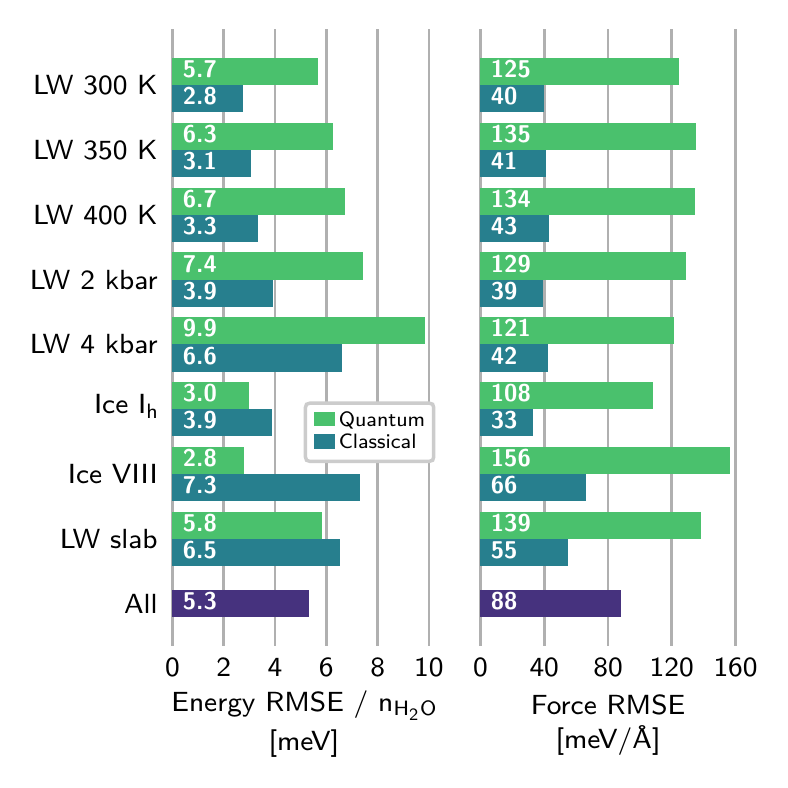}
\caption{\label{si_fig:test_set}
Root mean square error (RMSE) comparison for the generation 1 C-NNP model.
The bar chart shows RMSEs of energies and forces
for the C-NNP model with respect
to the revPBE0-D3 reference for an independently
generated test set.
The independent test set consists of 8000 uncorrelated
configurations covering the thermodynamic
conditions targeted during the development of
the C-NNP model and is decomposed into
the individual conditions of
liquid water (LW) at different temperatures and pressures,
hexagonal ice I$_\text{h}$ and ice VIII, as well as
the air-water interface (LW slab),
separately for a classical and quantum description of the nuclei.
See Sec.~\ref{sec:comp_det} of the main text for details on the creation of the independent test set.
}
\end{figure}

Next, we report the RMSE performance of
the generation 1 C-NNP model for the independently
generated test set in Fig.~\ref{si_fig:test_set}.
While the forces and energies of all
classical state points are reproduced with relatively
high accuracy, quantum configurations are not
reproduced as well by the generation 1 C-NNP model.
However, it is important to remember that
the associated training set consists of just
111 classical configurations of a single AIMD reference
trajectory.
In that respect, the performance is convincingly
good, especially for quantum structures.

\begin{figure}[!h]
\centering
\includegraphics{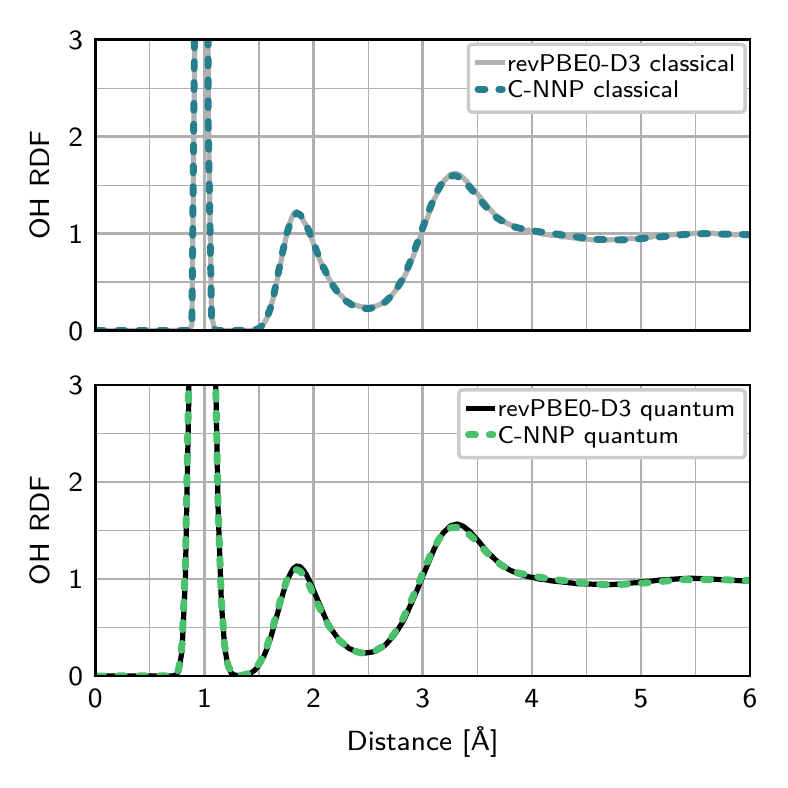}
\caption{\label{si_fig:rdfs-oh_gen1}
O-H RDFs complementary to the O-O RDFs shown in Fig.~\ref{si_fig:all_properties} for the generation 1 C-NNP model for both classical and quantum simulations of liquid water at 300~K.
}
\end{figure}

\begin{figure}[!h]
\centering
\includegraphics{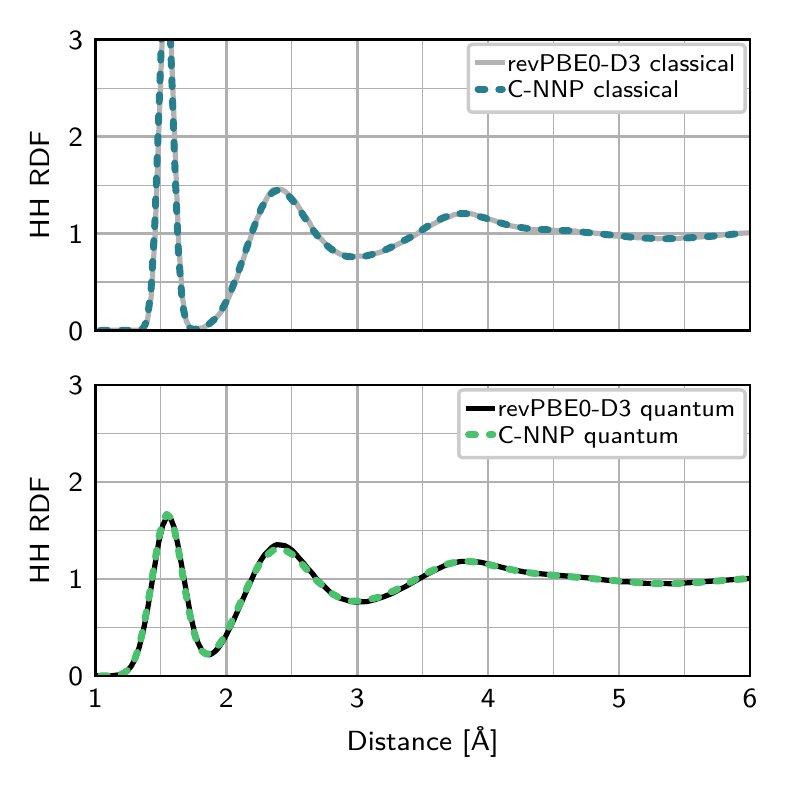}
\caption{\label{si_fig:rdfs-hh_gen1}
H-H RDFs complementary to the O-O RDFs shown in Fig.~\ref{si_fig:all_properties} for the generation 1 C-NNP model for both classical and quantum simulations of liquid water at 300~K.
}
\end{figure}

\begin{figure}[h!]
\centering
\includegraphics{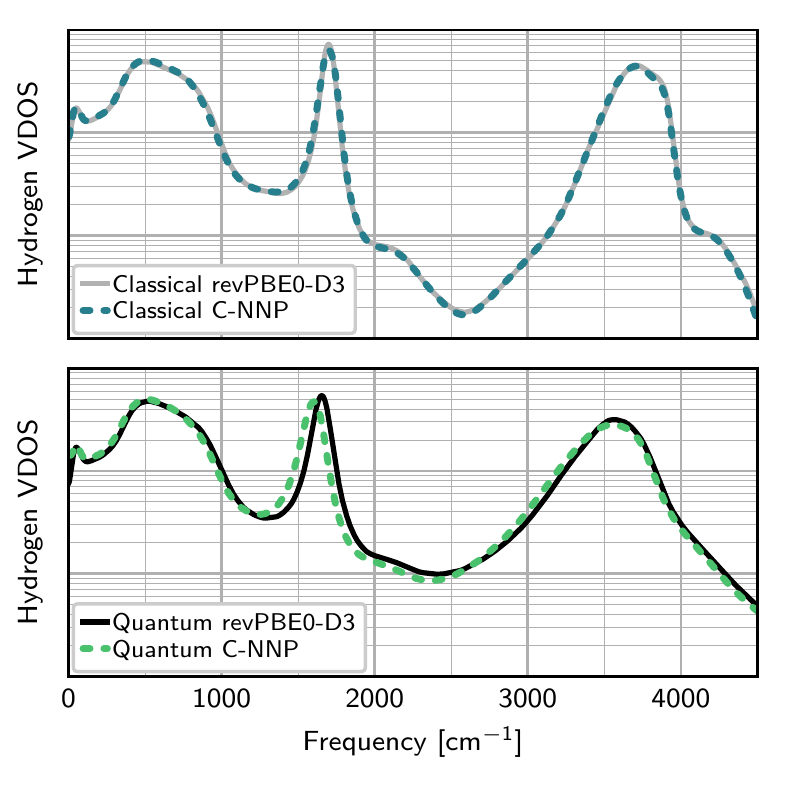}
\caption{\label{si_fig:vdos-logscale_gen1}
Hydrogen VDOS in logarithmic scale obtained from the generation 1 C-NNP model simulations.
}
\end{figure}

\begin{figure}[h!]
\centering
\includegraphics{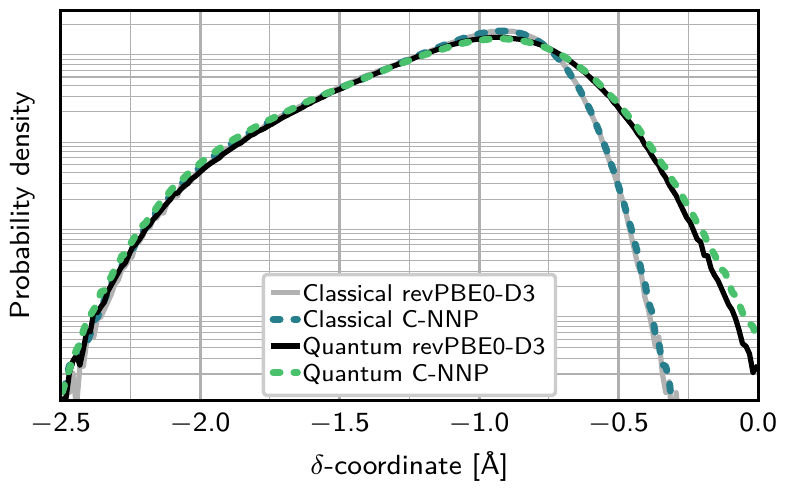}
\caption{\label{si_fig:delta_gen1}
Distributions of the proton-sharing coordinate $\delta$ for the generation 1 C-NNP model as obtained with classical and quantum nuclei for liquid water at 300~K, shown in logarithmic scale.
}
\end{figure}

Finally, we repeat the same detailed analysis
as presented in the previous section also for
the generation 1 C-NNP model
in Figs~\ref{si_fig:rdfs-oh_gen1}, \ref{si_fig:rdfs-hh_gen1}, \ref{si_fig:vdos-logscale_gen1}, and~\ref{si_fig:delta_gen1}.

\twocolumngrid

\FloatBarrier

\end{document}